\documentclass{jfm}
\usepackage{graphicx}
\usepackage{epstopdf, epsfig}

\usepackage{xcolor}

\usepackage{amsmath}
 \usepackage{amssymb}
 \usepackage{subcaption}
 \usepackage{makecell}
 \usepackage{blkarray}
 \usepackage{mathtools}
 \numberwithin{equation}{section}
\usepackage{stmaryrd}

\newcommand{\dx}{\mathrm{d}x}

\newcommand{\fp}[3]{\frac{\partial^{#3}#1}{\partial#2^{#3}}} 

\shorttitle{Surfactant cavity flow}
\shortauthor{Richard Mcnair,
  Oliver E. Jensen
 and Julien R. Landel}

\title{Surfactant spreading in a two-dimensional cavity and emergent contact-line singularities}

\author{Richard Mcnair\aff{1}\corresp{\email{richard.mcnair@postgrad.manchester.ac.uk}},
  Oliver E. Jensen\aff{1}\corresp{\email{oliver.jensen@manchester.ac.uk}}
 \and Julien R. Landel \aff{1}\corresp{\email{Julien.Landel@manchester.ac.uk}}}

\affiliation{\aff{1}Department of Mathematics, University of Manchester, Oxford Road, Manchester M13 9PL, UK}

\begin{document}

\maketitle

\begin{abstract}
We model the advective Marangoni spreading of insoluble surfactant at the free surface of a viscous fluid that is confined within a two-dimensional rectangular cavity.  Interfacial deflections are assumed small, with contact lines pinned to the walls of the cavity, and inertia is neglected.  Linearizing the surfactant transport equation about the equilibrium state allows a modal decomposition of the dynamics, with eigenvalues corresponding to decay rates of perturbations. Computation of the family of mutually orthogonal two-dimensional eigenfunctions reveals singular flow structures near each contact line, resulting in spatially oscillatory patterns of wall shear stress and a pressure field that diverges logarithmically.  These singularities at a stationary contact line are associated with dynamic compression of the surfactant monolayer. We  show how they can be regularized by weak surface diffusion. Their existence highlights the need for careful treatment in computations of unsteady advection-dominated surfactant transport in confined domains.
\end{abstract}


\section{Introduction}

Surfactants play a crucial role in a variety of natural and industrial flows \citep{kovalchuk2020}, including cleaning and decontamination \citep{landel2021}, transport in lung airways \citep{filoche2015, stetten2018}, and microfluidic applications \citep{temprano2018soap}.  Understanding how surfactants equilibrate near contact lines is also  important for flows over superhydrophobic surfaces, and slippery-liquid-infused-porous-surfaces (SLIPS) \citep{wang2020comparison}. As recently shown, surfactants can significantly affect the drag reducing properties of superhydrophobic surfaces when a flow concentrates them at stationary contact lines \citep{baier2021, landel2020, peaudecerf2017}. Surfactants are amphiphilic materials that accumulate at interfaces, where they lower surface tension.  Surfactant concentration gradients induce surface tension gradients, leading to so-called Marangoni flows of adjacent liquids that transport the surfactant itself \citep{manikantan2020}.  Here, we address a canonical surfactant transport problem in a two-dimensional confined geometry, showing how singular flow structures arise when spreading is impeded by a solid boundary.  

The self-induced spreading of a surfactant monolayer over a gas--liquid interface generates a variety of striking flow features \citep{afsar2003, matar2009dynamics, liu2019}.   If surface diffusion and solubility effects are weak, the leading edge of a localised surfactant monolayer spreading on an otherwise uncontaminated interface effectively rigidifies the interface locally.  For a monolayer spreading on a thin viscous film, this induces a jump in film depth via mass conservation effects \citep{dussaud2005, gaver1990dynamics}, that is captured within lubrication theory as a kinematic shock \citep{borgas1988monolayer}.  Lubrication theory also predicts a jump in surface shear stress, although a more refined analysis over shorter lengthscales reveals an integrable shear-stress singularity at the  leading edge of the monolayer, which can be regularised by surface diffusion or the presence of low levels of pre-existing (endogenous) surfactant \citep{jensen1998stress1}.   Out-of-plane displacement of the free surface may be suppressed by surface tension or gravity \citep{gaver1992droplet, jensen1992insoluble}. Inertial effects (which may be important if the initial spreading flow is sufficiently rapid) can generate an interfacial deflection at the  leading edge of the monolayer known as the Reynolds ridge, arising due to displacement effects in the viscous boundary layer beneath the spreading monolayer \citep{scott1982, jensen1998stress2}.   Spreading on thin films may also be accompanied by dramatic secondary fingering instabilities \citep{warner2004, jensen2006, liu2019}, showing the richness of surfactant flow phenomena.

The present paper addresses a complementary spreading problem, namely  surfactant spreading at the free surface of viscous fluid that is confined within a two-dimensional cavity.  We make a number of simplifying assumptions to aid our analysis: the free surface remains (almost) flat as a result of a restoring force, provided for example by surface tension, with contact lines pinned to the lateral walls of the cavity; the surfactant is insoluble and has a linear equation of state (a weakness discussed by \cite{swanson2015}); inertial effects are neglected and the Stokes flow is two-dimensional; molecular diffusion of surfactant at the free surface is  assumed negligible, except when analysing its impact on the regularisation of the singularities at the contact line; and the interface is pre-loaded with endogenous surfactant, to which exogenous surfactant is added. These modelling assumptions and their implications are discussed further in \S\ref{sec:disc}.

The aim of this study is to describe 
the interfacial and bulk transient flows produced by  self-induced Marangoni spreading of surfactant in a confined geometry. Exogenous surfactant added to the interface spreads, compressing the endogenous surfactant ahead of it \citep{grotberg1995, sauleda2021}.   Since the surfactant monolayer is laterally confined, the surfactant concentration rises at the pinned contact lines, while Marangoni stresses drive further surfactants towards the stationary boundary. Although there is no  motion of the contact line with respect to the solid wall, and therefore no risk of generating a non-integrable stress singularity associated with contact line motion \citep{huh71}, we nevertheless find that the unsteady Marangoni flow generates its own singular flow behaviour. We find two separate classes of singularities generated at the corner, one inducing an oscillatory shear stress pattern and the other a logarithmic pressure singularity.
In the main part of the paper, we focus our study to the case of a single-fluid flow with a free surface pinned at the contact line at an angle of $\pi/2$. This special case simplifies the numerical simulation and asymptotic analysis. In \S\ref{sec:disc} we relax these assumptions and discuss other wedge angles, between $0$ and $\pi$, and  two-fluid flows with arbitrary viscosities and where the surfactants lie at the interface. Our asymptotic analysis shows that the structure of the flow and the type of singularities identified for the single-fluid flow with $\pi/2$ contact-angle extend to this broader class of problems.

These findings complement studies of the lid-driven and shear-driven cavity problems \citep{shankar00},  benchmark problems for computational fluid dynamics, for which corner singularities are a recognised computational challenge \citep{kuhlmann2019}. Rather than tackle the full unsteady nonlinear problem numerically, our approach is to address the problem of small surfactant gradients, allowing the advective surfactant transport equation to be linearized.  When coupled to the linear Stokes flow in the cavity, we derive a problem that admits a decomposition into eigenmodes, with each eigenvalue representing the decay rate of a particular modal disturbance. This approach allows us to focus our numerical effort on capturing spatial structures.  While the temporal dynamics resembles a purely diffusive process, with mutually orthogonal modes decaying exponentially in time, each eigenmode has a singular form near the contact lines in the absence of surface diffusion.  We combine asymptotic and numerical approximations to obtain a full understanding of the flow structure at the interface and in the bulk.  

The model and the methods used to solve this surfactant Marangoni-driven cavity flow problem are described in \S\ref{sec:model}, with results presented in \S\ref{sec:results}.  Implications of the study are discussed in \S\ref{sec:disc}.




\section{Model}
\label{sec:model}

\begin{figure}
	\centering	\includegraphics[width=0.5\textwidth]{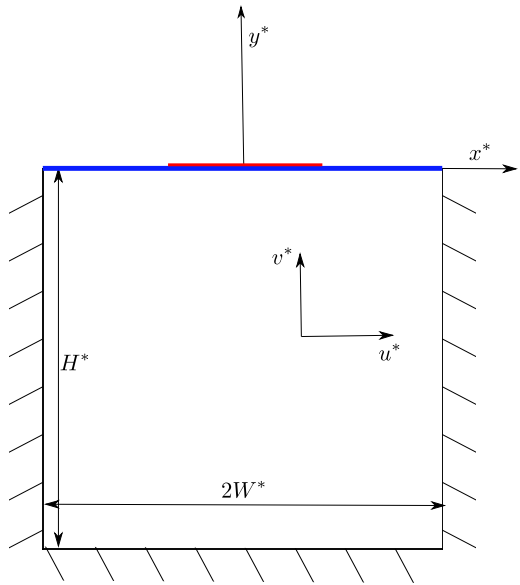}
	\caption{Diagram of the two-dimensional rectangular domain of the  flow problem. The flow is confined within rigid walls (hashed lines) and a free surface, for $-W^*\leq x^* \leq W^*$ and $-H^*\leq y^* \leq 0$. The incompressible Stokes flow in the bulk has velocity $u^*$ in the $x^*$ coordinate direction, and $v^*$ in the $y^*$ direction. At the free surface located at $y^*=0$ and $-W^*\leq x^* \leq W^*$, {an arbitrary initial non-uniform concentration profile of surfactant leads to an unsteady Marangoni flow that drives the flow in the bulk. The arbitrary initial concentration profile can be formed by  exogeneous surfactant (in red) deposited on the free surface at $t^*=0$, and some pre-existing endogenous surfactant with uniform concentration (in blue)}.}
	\label{fig:Domain}
\end{figure}


We model the spreading of an insoluble surfactant at the surface of an incompressible liquid of dynamic viscosity $\mu^*$ in a two-dimensional rectangular domain. The domain $V$ spans from $-W^* \leq x^* \leq W^* $ and $-H^*\leq y^* \leq 0$ with $x^*$ and $y^*$ in the horizontal and vertical directions, respectively, with $W^*$ the half width of the cavity,  $H^*$ its height, as shown in figure \ref{fig:Domain}.  Stars in superscript indicate dimensional variables. To model the flow induced by the surfactant spreading at the free surface, we  use the Stokes equations to relate the velocity field $\mathbf{u}^*(\mathbf{x}^*,t^*)=(u^*,v^*)$ to the pressure $p^*(\mathbf{x}^*,t^*)$, with $\mathbf{x}^*=(x^*,y^*)$ ignoring gravity and inertia. We impose the no slip and no penetration conditions on the three solid boundaries found at $x^*=-W^*$ and $W^*$ for $-H^*\leq y^*\leq0$ for the side walls, and $y^*=-H^*$ for $-W^*\leq x^*\leq W^*$ for the bottom wall. The free surface $\mathcal{F}$, assumed flat to leading order, is located at  $y^*=0$ for $-W^*\leq x^*\leq W^*$. 
We balance the  Cauchy stress $\boldsymbol{\sigma}^*\cdot \mathbf{n}$ at $\mathcal{F}$ (with the normal unit vector $\mathbf{n}=(0,1)$) with the gradient of the surface tension $\gamma^*$ tangentially and a restoring force due to strong surface tension normally.  The insoluble surfactant concentration $\Gamma^*$ at the surface is coupled to the flow via a time-dependent advective transport equation, and to the surface tension via an equation of state, assumed linear, which is valid for small variations of the concentration of surfactant from a reference concentration \citep{stone1990effects}.  
The governing equations are therefore
\begin{align}\label{eq:goveq}
		\boldsymbol{\nabla}^* p^* &=  
		\mu^*\nabla^{*2} \mathbf{u}^*, \quad 
		\boldsymbol{\nabla}^*\cdot \mathbf{u}^* = 0, \quad
	\Gamma^*_{t^*} = - \frac{\partial}{\partial x^*} 
	\left(u_s^*\Gamma^* \right), \quad
	\gamma^* = \gamma_0^* - (\gamma_0^* - \gamma_c^*) \frac{\Gamma^*}{\Gamma_c^*},
\end{align}
where subscript $s$ denotes evaluation on $\mathcal{F}$, $\gamma_0^*$ is the reference surface tension and $\gamma_c^*$ is its (lower) value when the surfactant is at a reference concentration $\Gamma_c^*$. 
The  boundary conditions associated with (\ref{eq:goveq}) are
\begin{subequations}
\begin{align}
	\mathbf{u}^* &= \mathbf{0} \quad \text{ on } x^*=\pm W^*, \quad y^*=-H^*,
	\\
	\mathbf{u}^* \cdot \mathbf{n} &= 0 \quad \text{ on } y^*=0,
	\\
		\boldsymbol{\sigma}^* \cdot \mathbf{n} &= -\gamma^* \mathbf{n} (\boldsymbol{\nabla}^*\cdot \mathbf{n}) + \frac{\partial \gamma^*}{\partial x^*} 
		\mathbf{t} \quad \text{on} \quad \mathcal{F},
\end{align}
\label{eq:bcs}
\end{subequations}
with $\mathbf{t}=(1,0)$ the tangential unit vector at the free surface. We have made the assumption that $S^*= \gamma_0^*-\gamma_c^*\ll \gamma_c^*$, so that the  surface tension remains sufficiently large, in comparison to its reduction by surfactant $S^*$, to suppress deflections of the free surface from $y^*=0$. Effectively, the  capillary number in our problem is $S^*/\gamma_c\ll 1$, since $S^*$ is a characteristic viscosity--velocity scale. This small capillary number assumption is discussed further in \S\ref{sec:results} and \ref{sec:disc}.  Hence, the leading-order kinematic boundary condition reduces to (\ref{eq:bcs}b) and any surface curvature can be neglected, such that $W^*\boldsymbol{\nabla}^*\cdot \mathbf{n}\ll 1$. We will exploit the normal stress condition later to evaluate the small surface deflections induced by the flow, while imposing the tangential component of (\ref{eq:bcs}\textit{c}) on $\mathcal{F}$.


Using the length scale $W^*$, velocity scale $S^*/\mu^*$ and pressure scale $S^*/W^*$, we relate dimensional starred variables to their dimensionless counterparts via 
\begin{subequations}
	\begin{equation}
\mathbf{x}= (x,y)= \left(\frac{x^*}{W^*},\frac{y^*}{W^*} \right), \quad  H=\frac{H^*}{W^*}, \quad \Gamma = \frac{\Gamma^*}{\Gamma_c^*}, \quad \gamma = \frac{\gamma^*-\gamma_c^*}{S^*},
	\end{equation}
	\begin{equation}
\mathbf{u}=(u,v)=\frac{\mu^*  }{ S^*}(u^*,v^*), \quad t = \frac{S^* t^* }{W^* \mu^*}, \quad p = \frac{ W^* p^*}{S^*}.
	\end{equation}
\label{scales}
\end{subequations}
After rescaling, the governing equations become, in the bulk,
\begin{subequations}
\begin{equation}
	\begin{pmatrix}
		p_x \\ p_y
	\end{pmatrix} = 
	\begin{pmatrix}
		u_{xx} +u_{yy} \\ 
		v_{xx} +w_{yy}
	\end{pmatrix}, \quad
  u_x+v_y = 0 ,\quad 
\label{incompessibility}
	\gamma = 1 - \Gamma,
\end{equation} 
with, at the free surface,
\begin{equation}
		{ \Gamma}_{ t} = - (\Gamma u)_x , \quad
 u_y = - \Gamma_x, \quad
v = 0, \quad \text{ on } y = 0
 \label{B13}
\end{equation}
and
\begin{equation}
\label{B12}
	\mathbf{u} = \mathbf{0}, \quad \text{ on } x=\pm1, \text{ and }  y = -H.
\end{equation}
\end{subequations}
We introduce a stream function $\psi(x,y,t)$ such that $\psi_y=u$, and $\psi_x=-v$, enforcing incompressibility.  The problem reduces to the biharmonic equation in the bulk
%
\begin{equation}
	\nabla^4 \psi = 0
	\label{eq:bih}
\end{equation} 
subject to
\begin{subequations} 
\label{UnIntegratedBoundaryCond}
\begin{gather}
\Gamma_t=-(\Gamma \psi_y)_x,\quad	 \psi_{yy} = -\Gamma_{x}, \quad \psi=0 \quad \text{on} \quad y=0, \label{UnIntegratedBoundaryConda} \\
	\psi_x(\pm 1,y) = \psi(\pm 1,y) = 0, \\ 
	\psi(x,-H)=\psi_y(x,-H) = 0.
\end{gather} 
\end{subequations}
The stream function vanishes at the four boundaries of the domain, in order to impose the no-flux boundary condition. The problem is closed by imposing an initial surfactant profile, representing the addition of exogenous surfactant to an  endogenous monolayer initially present on the interface. We note that the transport equation (\ref{UnIntegratedBoundaryCond}a) is linear in $\Gamma$, given a surface velocity $\psi_y$.  Therefore writing $\Gamma$ as the sum of endogenous and exogenous components, $\Gamma=\Gamma_1+\Gamma_2$ say, the two components satisfy the same transport equation $\Gamma_{it}=-(\Gamma_i\psi_y)_x$ ($i=1,2$), allowing the evolution of the components to be tracked individually if necessary \citep{grotberg1995}. From (\ref{UnIntegratedBoundaryCond}\textit{a,b}) we anticipate the presence of singularities at the contact lines $(x,y)=(\pm 1,0)$, as the boundary conditions are discontinuous here. We discuss these in detail in \S\ref{sec:corners} below, to understand their impact on the surface and bulk velocity fields and the surfactant distribution.
 
\subsection{Linearization}\label{sec:linearization}

At large times, the surfactant relaxes to a uniform level $\bar{\Gamma}$ and the velocity field decays to zero.  We perturb the system around this state, noting that the resulting problem is homogeneous.  We decompose the solution into individual eigenmodes, writing for one such mode 
\begin{equation}
	\Gamma(x,t) = \bar{\Gamma} + \hat{\Gamma}(x)e^{-\lambda t}, \quad \mathrm{where}\quad \int_{-1}^1 \hat{\Gamma}(x)\,\mathrm{d}x=0,
	\label{eq:larget}
\end{equation} 
assuming the same time-dependence for other variables, for example $u(x,y,t) = \hat{u}(x,y)e^{-\lambda t}$.  The surfactant transport equation at the free surface (first equation in \eqref{UnIntegratedBoundaryConda}) is the only equation that changes under linearization, becoming
\begin{equation}
\alpha \hat{\Gamma} = \hat{u}_{x}, \quad \text{on } y=0,
\label{eq:bc4}
\end{equation}
where $\alpha = \lambda/\bar{\Gamma}$. Equation (\ref{eq:bc4}) is valid for small surfactant concentration perturbation, or at large times since we assume $\vert \hat{\Gamma}(x)\vert e^{-\lambda t}\ll \bar{\Gamma}$.
Equation (\ref{eq:bc4}) may be combined with 
the stress condition $\Gamma_x = - u_y$ to give the homogeneous boundary condition
\begin{equation}
	-\alpha \hat{u}_{y} = \hat{u}_{xx}, \quad \text{on } y=0.
\end{equation}
The resulting eigenvalue problem for the perturbation stream function may then be stated as the biharmonic equation (\ref{eq:bih}), under the boundary conditions (\ref{UnIntegratedBoundaryCond}\textit{b,c}) for $\hat{\psi}$ and
\begin{equation}
	\alpha \hat{\psi}_{yy} = -\hat{\psi}_{xxy} \qquad \hat{\psi}=0, \quad \text{on } y=0.
	\label{eq:bc3}
\end{equation}
We seek the set of decay rates  $\alpha_n$, $n=1,2,\dots$, which are  eigenvalues for which a solution exists with the corresponding eigenmodes $\hat{\psi}_n$. We distinguish two families for the decay rates $\alpha_n$ and eigenmodes $\hat{\psi}_n$, such that $\hat{\psi}_n$ is either an even or odd function of $x$. For the rest of this study the subscript $n$ will refer to the odd modes  unless otherwise specified. From each of these eigenmodes and eigenvalues we can derive the associated surfactant distribution $\hat{\Gamma}_n$, the shear stress at the free surface $\tau_n(x) = \partial \hat{u}_{n}(x,0)/\partial y $, the vorticity field $\hat{\omega}_n = -\nabla^2\hat{\psi}_n$, and the pressure field $\hat{p}_n$, obtained from the vorticity via $\nabla \hat{p}_n = (-\partial \hat{\omega}_n/\partial y,\partial \hat{\omega}_n/\partial x)$.  

This problem has two key global characteristics.  An energy dissipation argument (see Appendix~\ref{app:energy}) shows that all the decay rates satisfy
\begin{equation}
    \alpha_n=\frac{\int_V \hat{\omega}_n^2 \,\mathrm{d}A}{\int_{-1}^1 \hat{\Gamma}_n^2 \, \mathrm{d}x},
    \label{eq:energy}
\end{equation}
where $V$ is the rectangular flow domain (see figure~\ref{fig:Domain}).  Application of the reciprocal theorem \citep{masoud_stone_2019} 
(see Appendix~\ref{app:ortho}) yields the orthogonality condition
\begin{equation}\label{eq:orthcond}
    \int_{-1}^1 \hat{\Gamma}_m \hat{\Gamma}_n \,\mathrm{d}x=0, \quad  \forall \ m\neq n.
\end{equation}
When solving an initial value problem with time-dependent surfactant and velocity fields, the condition (\ref{eq:orthcond}) can be used to project the initial condition for $\hat{\Gamma}$ onto its component modes. The initial Marangoni gradient that drives the flow can arise for example from the addition of exogenous surfactant to an otherwise uniform endogenous surfactant distribution. In this study, we do not track the interface between these distributions explicitly, and we assume that the exogenous and endogenous surfactant concentrations add together to form a single concentration field \citep{grotberg1995}.

\subsection{Finite-difference numerical solution}
\label{subsec:finitedifference}

We compute a numerical solution $\widetilde{\psi}_n$ (where a wide tilde denotes a numerically computed variable)  of the unknown perturbation modes of the stream function, $\hat{\psi}_n$, satisfying the biharmonic equation (\ref{eq:bih}) and the boundary conditions (\ref{UnIntegratedBoundaryCond}\textit{b,c}) and (\ref{eq:bc3}), using a finite-difference scheme. Using a row$\times$column ordering convention, the domain described in figure~\ref{fig:Domain} is discretized as an $M\times N$ rectangular grid, in the $y$ and $x$ directions, respectively, with uniform spacing $\Delta y$ and $\Delta x$, and supplemented with a set of ghost-points around the periphery creating an $(M+2)\times(N+2)$ grid. We use a second-order-accurate 13 point stencil, involving standard finite-difference approximations of derivatives \cite[see e.g.][]{fornberg1988generation}, to approximate the biharmonic operator in \eqref{eq:bih} on the grid of unknowns. These unknowns correspond to the unknown values of the perturbation stream function of a given mode at a given grid point $\widetilde{\psi}_n(x_j,y_i)$, with $3\leq i\leq M$, $3\leq j\leq N$. 
We impose $\widetilde{\psi}_n(x_j,y_i)=0$ at the boundaries of the domain $j=2$ and $j=N+1$, with $2\leq i\leq M+1$, and $i=2$ and $i=M+1$, with $2\leq j\leq N+1$, to implement the boundary conditions in (\ref{UnIntegratedBoundaryCond}\textit{b,c}) and (\ref{eq:bc3}) which state that the  stream function vanishes at all the boundaries. The $(M-2)\times(N-2)$ grid of unknowns is expressed as a column vector $\boldsymbol{\psi}$ which is assembled by the vertical concatenation of the rows of unknowns of $\widetilde{\psi}_n(x_j,y_i)$. The numerical operator modelling the biharmonic operator on the grid can then be approximated by a matrix operating on $\boldsymbol{\psi}$. The remaining no-slip boundary conditions in  (\ref{UnIntegratedBoundaryCond}\textit{b,c}) at the walls and the surfactant boundary condition in (\ref{eq:bc3}) at the free-surface can be approximated by a finite difference discretisation acting on the $M\times N$ grid and the ghost points. Values of the stream function at the ghost points are calculated as functions of the values in the interior points and added to the linear system such that the boundary conditions are satisfied. The system can then be rearranged so that $\boldsymbol{\psi}$ satisfies,
\begin{equation}
	\mathsf{B} \boldsymbol{\psi}= \alpha\mathsf{C}\boldsymbol{\psi},
	\label{eq:fd}
\end{equation}
where $\mathsf{B}$ and $\mathsf{C}$ are sparse $(N-2)(M-2)\times (N-2)(M-2)$ matrices. Equation \eqref{eq:fd} represents a generalised numerical eigenvalue problem for the eigenmodes $\widetilde{\psi}_n$ and the associated numerically-calculated decay rates $\widetilde{\alpha}_n$, from the free-surface boundary condition (\ref{eq:bc3}). We solved \eqref{eq:fd} using the MATLAB function \texttt{eigs}.  From the solutions for each mode $\widetilde{\psi}_n$  we compute numerical approximations of all other quantities of interest for each mode, such as the surfactant concentration profile $\widetilde{\Gamma}_n(x)$, the surface stress $\widetilde{\tau}_n(x) = \widetilde{\psi}_{n,yy}(x,y=0)$, the velocity field $(\widetilde{u}_n,\widetilde{v}_n)(x,y)$, the vorticity field $\widetilde{\omega}_n(x,y)$, and the pressure field $\widetilde{p}_n(x,y)$. The solutions $\widetilde{\psi}_n$ are normalised by requiring $\max(\widetilde{\Gamma}_n(x)) - \min(\widetilde{\Gamma}_n(x)) = 1$ for each $n$.

\begin{table}
    \centering
    \begin{tabular}{c|c|c|c}
    \hline
   Even modes & Eigenvalue $\widetilde{\alpha}_n$ & Eigenvalue $\breve{\alpha}_n$ from \eqref{eq:energy} & Relative error $(\widetilde{\alpha}_n-\breve{\alpha}_n)/\widetilde{\alpha}_n$ \\ \hline
      1 &  0.57313 & 0.57290 & 0.00039  \\
      2 &  2.10723 & 2.10738 & -0.00007  \\
      3 &  3.67670 & 3.67710 & -0.00011  \\
      4 &  5.24715 & 5.24820 & -0.00020  \\
      \hline
      Odd modes 
      \\ \hline
      1 &  1.29728 & 1.29725 & 0.00002  \\
      2 &  2.88997 & 2.89018 & -0.00007  \\
      3 &  4.46164 & 4.46232 & -0.00015  \\
      4 &  6.03235 & 6.03390 & -0.00026  \\
      \hline
    \end{tabular}
    \caption{Decay rates predicted as eigenvalues $\widetilde{\alpha}_n$ computed numerically from (\ref{eq:fd}) compared to $\breve{\alpha}_n$, which are those computed from eigenmodes using (\ref{eq:energy}) for $H=2$, found using a $4000\times4000$ grid. The relative error provides a measure of global numerical error, suggesting that the values for $\widetilde{\alpha}_n$ are accurate up to three significant figures.}
\label{tab:decayrates}
\end{table}

We use global integrated measures to estimate the accuracy of the computational scheme, such as the energy balance (\ref{eq:energy}), see table~\ref{tab:decayrates}, and mode orthogonality, see Appendix \ref{app:ortho}. The  relative error of the decay rates computed directly as eigenvalues from \eqref{eq:fd} and indirectly from the eigenmodes via \eqref{eq:energy} (denoted as $\breve{\alpha}_n$) remains less than $4\times10^{-4}$ for all odd and even modes calculated (up to $n=4$) for $H=2$ using a $4000\times4000$ grid (see table~\ref{tab:decayrates}). For the same refinement and the same modes, the concentration profiles $\widetilde{\Gamma}_n$ are orthogonal with an absolute error less than $8\times10^{-6}$ (see \eqref{eq:ortho}). We use asymptotic methods, described below, to assess and contain (via global grid refinement) the inevitable local numerical inaccuracies associated with corner singularities at the contact lines $(x,y)=(\pm 1,0)$. 

\subsection{Corner asymptotics}
\label{sec:corners}

We anticipate from the outset the appearance of Moffatt vortices \citep{moffatt1964viscous} in the lower corners of the domain, at $(x,y) = (\pm 1,-H)$, as we will show in our results presented in \S\ref{sec:results}.  However, the structure of the flow 
at the top corners of the domain, where the surfactant-laden surface meets the wall, needs separate asymptotic treatment. 
We illustrate this at the top left corner, introducing polar coordinates $(r,\theta)$ centred on $(x,y)=(-1,0)$ with $\theta=0$ along $y=0$ (and $r$ increasing in the positive $x$ direction) and $\theta=-\pi/2$ along $x=-1$ (and $r$ increasing in the negative $y$ direction) and  
seeking separable solutions of the form
\begin{equation}
	\hat{\psi}(r,\theta) \approx \Re\left[ \sum_i A_i r^{\Phi_i} f_{\Phi_i}(\theta) \right], \quad \text{for} \ r\rightarrow 0.
	\label{eq:psiexp}
\end{equation} 
Here $\Re\left[\cdot\right]$ indicates taking the real part, noting that the amplitudes $A_i$ and exponents $\Phi_i$ of local modes of the biharmonic equation may be complex. We will find that the sum in (\ref{eq:psiexp}) is a sum of multiple countable series. For notational simplicity, we will use $\Phi$ to represent exponents, and we will suppress the subscript $n$ on $\hat{\psi}$ and $\alpha$ associated with each eigenmode. To satisfy the governing biharmonic equation (\ref{eq:bih}),  $\nabla^4\hat{\psi}(r,\theta)=0$, and the boundary conditions (\ref{UnIntegratedBoundaryCond}\textit{b,c}), $\hat{\psi}(r,0) = \hat{\psi}(r,-\pi/2) = \hat{\psi}_\theta(r,-\pi/2)=0$, 
starting from more general formulas for the $\theta$-dependent functions \citep{moffatt1964viscous}, we find that
\begin{subequations}
\begin{align}\label{f_1}
	f_1(\theta) &= 
	\frac{\pi}{2}\sin{\theta} - \frac{2}{\pi}\theta \cos{\theta} +\theta \sin{\theta} 
	 & (\Phi=1),\\
\label{f_2}
	f_2(\theta) &= 
	\cos{(2\theta)}-\frac{2}{\pi}\sin{(2\theta)} - \frac{4 \theta}{\pi} -1 
     & (\Phi=2),
\end{align}
and generally for $\Phi > 0$, $\Phi \neq 1,2$,
\begin{multline}\label{f_mu}
	f_{\Phi}(\theta) =
	\frac{\cos{(\Phi\pi/2)}\sin{(\Phi \pi/2)}}{\sin^2{(\Phi\pi/2)}-\Phi } (\cos{(\Phi\theta)}-\cos{((\Phi-2)\theta)})\\ + 
	\frac{\cos^2{(\Phi\pi/2)}-\Phi+1}{\sin^2{(\Phi\pi/2)}-\Phi }\sin{(\Phi\theta)}  +\sin{((\Phi-2)\theta)},
\end{multline} 
which in the special case of integer $\Phi$ reduces to
\begin{equation}\label{f_mu odd}
	f_{\Phi}(\theta) = 
\begin{cases}
	\sin{(\Phi\theta)}+\sin{((\Phi-2)\theta)}, 
&  \quad $if $\Phi$ is an odd integer strictly greater than $1$,$ \\
\frac{\Phi-2}{\Phi} \sin(\Phi\theta)+\sin((\Phi-2)\theta),
& \quad $if $\Phi$ is an even integer strictly greater than $2$.$ 
\end{cases}
\end{equation}  
\end{subequations}
%
The final boundary condition, the surfactant boundary condition at the free surface in (\ref{eq:bc3}): $-\alpha \hat{\psi}_{yy} = \hat{\psi}_{xxy}$ at $y=0$, is, in polar coordinates,
\begin{equation}
	-\alpha\left( \frac{1}{r^2} \hat{\psi}_{\theta \theta} + \frac{1}{r} \hat{\psi}_{r}\right)
	=\frac{1}{r} \hat{\psi}_{rr \theta} -\frac{2}{r^2}\hat{\psi}_{r \theta} + \frac{2}{r^3} \hat{\psi}_{\theta}, \quad \text{for} \ \theta=0.
	\label{eq:topbc}
\end{equation} 
Writing the first few terms in the expansion (\ref{eq:psiexp}) for $\hat{\psi}$ in the form
\begin{equation}\label{AsympForm}
	\hat{\psi}  = \Re \left[ A_a r^{a}f_a(\theta)+A_b r^bf_b(\theta)+ A_c r^cf_c(\theta)+\dots \right]
\end{equation} 
where $a$, $b$, $c$... are complex numbers (representing exponents $\Phi_i$) such that ${\Re}(a)<{\Re}(b)<{\Re}(c)$ and $A_a\neq 0$, 
we obtain from (\ref{eq:topbc}) (after 
multiplying by $r^3$)
\begin{multline}\label{AsympRelation}
	-\alpha ( A_a r^{a+1}f^{\prime \prime}_a(0) + A_b r^{b+1}f^{\prime \prime}_b(0) + A_c r^{c+1}f^{\prime \prime}_c(0) + \dots ) 
	\\
	= \left(a^2-3a +2\right)A_a r^{a}f_a^{\prime}(0) + \left(b^2-3b +2\right)A_b r^{b}f_b^{\prime}(0) + \left(c^2-3c +2\right)A_c r^{c}f_c^{\prime}(0) + \dots
\end{multline} 
The $r^a$ term is dominant as $r \to 0$, imposing that
\begin{equation}\label{eq:fcondition}
	(a-1)(a-2)f_a^{\prime}(0)=0.
\end{equation} 
This equation opens multiple possibilities. 
The first case $a=1$ corresponds to a solution with non-integrable stress $\tau(r) = \hat{\psi}_{\theta \theta}(\theta=0)/r^2$, and therefore unbounded surfactant concentration as $r\rightarrow 0$, so we impose $A_1=0$.  This leaves two other cases: $a=2$ and $f^{\prime}_a(0)=0$.  The latter third case
yields an infinite set of complex exponents of the type described by  \cite{moffatt1964viscous}, each representing a homogeneous local solution, which we will analyse further below.  The expansion (\ref{eq:psiexp}) therefore constitutes multiple independent series.  


In the second case, with $a=2$, \eqref{AsympRelation} becomes
\begin{multline}\label{eq:Case2balance}
    	-\alpha (A_2 r^{3}f^{\prime \prime}_2(0) + A_b r^{b+1}f^{\prime \prime}_b(0) + A_c r^{c+1}f^{\prime \prime}_c(0) + \dots ) = \left(b^2-3b +2\right)A_b r^{b}f_b^{\prime}(0)\\ + \left(c^2-3c +2\right)A_c r^{c}f_c^{\prime}(0) + \dots
\end{multline} 
The dominant balance  must be between the $r^b$ and $r^3$ terms since we imposed ${\Re}(a)<{\Re}(b)$, hence $b=3$. We then have
\begin{equation}\label{eq:A2A3}
	-\alpha A_2 f_2^{\prime \prime}(0) = 2A_3 f_3^{\prime}(0).
\end{equation} 
Using \eqref{f_2} and \eqref{f_mu odd} for $f_2$ and $f_3$, respectively, (\ref{eq:A2A3}) becomes
$\alpha A_2 = 2A_3$.
The next balance in \eqref{eq:Case2balance} gives 
$-\alpha A_3 f_3^{\prime \prime}(0) = 6A_4 f_4^{\prime}(0)$,
however from \eqref{f_mu odd} we can see that $f_3^{\prime \prime}(0)=0$, which implies that $A_4=0$ and  this series terminates. The first series contributing to $\hat{\psi}$ is therefore simply
\begin{multline}\label{eq:realseriesPsi}
	 A_2r^2 \left(f_2(\theta) +\frac{\alpha}{2} r f_3{(\theta)} \right)
	= \\
	A_2r^2\left(\cos{(2\theta)}-\frac{2}{\pi}\sin{(2\theta)} - \frac{4 \theta}{\pi} - 1 +\frac{\alpha}{2} r(\sin{(3\theta)}+\sin{\theta})\right).
\end{multline}

In the third case, setting $f_a^{\prime}(0)=0$ in \eqref{eq:fcondition} (with $a\neq 1$ and $2$) 
gives 
\begin{equation}
	\frac{\cos^2{(a\pi/2)}-a+1}{\sin^2{(a\pi/2)}-a }a  +(a-2) = 0,
\end{equation} 
so that the complex roots satisfy
\begin{equation}
  \sin^2{(a\pi/2)} -a(2-a)=0.
	\label{eq:g}
\end{equation}  
We label the roots of (\ref{eq:g}) which have non-zero imaginary part as $a_1, a_2,\dots$ with $0<\Re (a_1)<\Re (a_2)<\dots$. The roots $a_i$ are independent of $H$, and correspond to the exponents in Moffatt's series \citep{moffatt1964viscous} for anti-symmetric Stokes flow in a right-angle corner subject to arbitrary disturbance at a large distance. The first five complex roots are shown in table~\ref{tab:1}.
\begin{table}
    \centering
    \begin{tabular}{l|l}
         $a_1 \approx 3.7396 + 1.1190 i \quad $ & $\quad K_{11} \approx -0.036666 + 0.076161 i$\\
         $a_2  \approx 5.8083 + 1.4639 i$  & $\quad K_{12} \approx 0.046960 + 0.002377 i$\\
         $a_3 \approx 7.8451 + 1.6816 i$ & $\quad K_{13} \approx -0.001195 + 0.000349 i$\\
         $a_4 \approx 9.8688 + 1.8424 i$ & $\quad K_{14} \approx 0.000198 + 0.000122 i$\\
         $a_5 \approx 11.886 + 1.9702 i$ & $\quad K_{15} \approx -0.000002 - 0.000002 i$
    \end{tabular}
    \caption{Left: Approximate numerical values of the first five complex roots $a_i$ of equation (\ref{eq:g}). These roots are the exponents in Moffatt's series for anti-symmetric Stokes flow \citep{moffatt1964viscous} subject to a disturbance at a large distance in the case of right-angle corners.  Right: approximate values of the first five coefficients $K_{ij}$ for the dominant root $i=1$ in the series expansion (\ref{eq:kseries}) associated with $\hat{\psi}$ and computed using \eqref{eq:Kij}. All of these quantities are independent of the global parameter $H$.}
    \label{tab:1}
\end{table}
Each complex root $a$ generates its own asymptotic series of the form \eqref{AsympForm} with $a=a_i$, $b=b_i$, $c=c_i$, etc. where $0<\Re (a_i)<\Re (b_i)<\dots$, for $i=1,2,3,\dots$ For example, for $i=1$ 
 \eqref{AsympRelation} gives
\begin{multline}
	-\alpha ( A_{a_1} r^{a_1+1
	}f^{\prime \prime}_{a_1
	}(0) + A_{b_1} r^{b_1+1}f^{\prime \prime}_{b_1}(0) + A_{c_1} r^{c_1+1}f^{\prime \prime}_{c_1}(0) + \dots ) =   
	\\ \left(b_1^2-3b_1 +2\right)A_{b_1} r^{b_1}f_{b_1}^{\prime}(0) + \left(c_1^2-3c_1 +2\right)A_{c_1} r^{c_1}f_{c_1}^{\prime}(0) + \dots
\end{multline} 
such that the dominant balance is $b_1=a_1+1
$, requiring that
\begin{equation}
	-\alpha A_{a_1}	f^{\prime \prime}_{a_1 
	}(0)  = a_1(a_1-1)A_{b_1} f_{b_1 
	}^{\prime}(0),
\end{equation} 
with the approximate value of $a_1$ given in table~\ref{tab:1}.
From such relations for $a_1$, $b_1$, $c_1,\dots$ we can derive the associated contribution to $\hat{\psi}$, of the form
\begin{equation}
A_{a_1} \left( r^{a_1}f_{a_{1}}(\theta) +\alpha K_{11}  r^{a_1+1} f_{a_1+1}(\theta) + \alpha^2 K_{12} r^{a_1+2}f_{a_1+2}(\theta) + \dots \right),
\label{eq:kseries}
\end{equation} 
where the first five coefficients $K_{1j}$ related to $a_1$ in (\ref{eq:kseries}) are shown in table~\ref{tab:1}. These coefficients can be computed through the recurrence relation

\begin{equation}\label{eq:Kij}
    K_{ij} = -\frac{K_{i(j-1)} f^{\prime \prime}_{a_i+j-1}(0)}{((a_i+j)^2-3(a_i+j)+2)f^{\prime }_{a_i+j}(0)},
\end{equation}
for integers $i\geq 1$ and $j\geq 1$ and with $K_{i0}=1$.

In summary, the full asymptotic series for the $n$th eigenmode in the neighbourhood of the contact line, which we originally stated in the form  \eqref{eq:psiexp}, is  
\begin{equation}
\hat{\psi}_n(r,\theta)=
A_{n2}r^2 \left[f_2(\theta) +\frac{\alpha_n}{2} r f_3{(\theta)} \right] + \Re\left[\sum_{i=1}^{\infty}A_{na_i} \left(\sum_{j=0}^\infty \alpha_n^j K_{ij} r^{a_i+j}f_{a_i+j}(\theta) \right)
\right].
\label{eq:psifull}
\end{equation} 
The coefficients $K_{ij}$ can be computed from the recurrence relation (\ref{eq:Kij}), whilst the coefficients $A_{n2}$ and $A_{na_i}$ must be determined from fitting to numerical solutions. Analysing (\ref{eq:psifull}) we can notice that $\hat{\psi}_n$ approaches different values as $r\to0$ for different values of $\theta$, capturing the stream function's singular behaviour in this limit. Away from the corner, the complex exponents involved in the second term of (\ref{eq:psifull}) produce oscillatory behaviour as a function of $r$, such that for the corresponding parameters, (\ref{eq:psifull}) is capable of producing Moffat-type eddies at the top corners, which are observed in the numerical solution for higher-order modes (shown in \S\ref{sec:results}). This is a local expansion at the top corner, however global information about the aspect ratio of the domain and the global symmetry or anti-symmetry of $\hat{\psi}_n$ enters into this expression through $\alpha_n$. 
We note for later reference that the surface shear stress $\tau_n(x)$ has the local form
\begin{equation}
    \hat{\tau}_n(x) = \left.\fp{\hat{u}_{n}}{y}{}\right|_{y=0} \approx -4A_{n2} + \Re \left\{ A_{na_1} (1+x)^{a_1-2} f_{a1}''(0) +\dots \right\} \quad\mathrm{as} \quad x\rightarrow -1,
    \label{eq:shearwall}
\end{equation}
which has a non-zero asymptotic value at the contact lines $\hat{\tau}_n(-1)= - 4A_{n2}$. Moreover, the complex roots $a_i$ imply that the value of the shear stress oscillates as $x\to -1$. Near the contact line $(x,y)=(-1,0)$, the leading-order surfactant distribution has a finite non-zero gradient, computed from (\ref{eq:bc4}), 
\begin{equation}
    \hat{\Gamma}_n(x) \approx -\frac{8A_{n2}}{\pi\alpha_n}+4A_{n2}(1+x)+\dots \quad\mathrm{as} \quad x\rightarrow -1.
     \label{eq:Gamma}
\end{equation}
The leading-order vorticity near the contact line $(x,y)=(-1,0)$ has the form
\begin{equation}
\hat{\omega}_n\approx 4A_{n2}\left(1+\frac{4}{\pi}\tan^{-1}\frac{y}{1+x}\right)+\dots, 
    \label{eq:vortcorner}
\end{equation}
which shows that the vorticity is multi-valued at the corner owing to the singularity and depending on the angle of approach. From the vorticity, we find that the leading-order pressure locally is
\begin{equation}
\hat{p}_n\approx -\frac{8A_{n2}}{\pi}\log(y^2+(1+x)^2)+\dots,
\label{eq:presswall}
\end{equation}
which shows that the pressure diverges in a logarithmic fashion at the corner. The above expressions give the local expansion at the top left corner of the domain. A similar expansion is then trivially obtained at the top right corner by symmetry. These asymptotic results complement the numerical solutions, less accurate near the contact lines $(x,y)=(\pm1,0)$, providing a complete understanding of the effect of the corner singularities on the relevant physical quantities in this problem.
  
\section{Results}
\label{sec:results}

\begin{figure}
	\centering
	\begin{subfigure}{0.495\textwidth}
		\includegraphics[trim={0cm 0cm 1cm 0cm},clip,width=\textwidth]{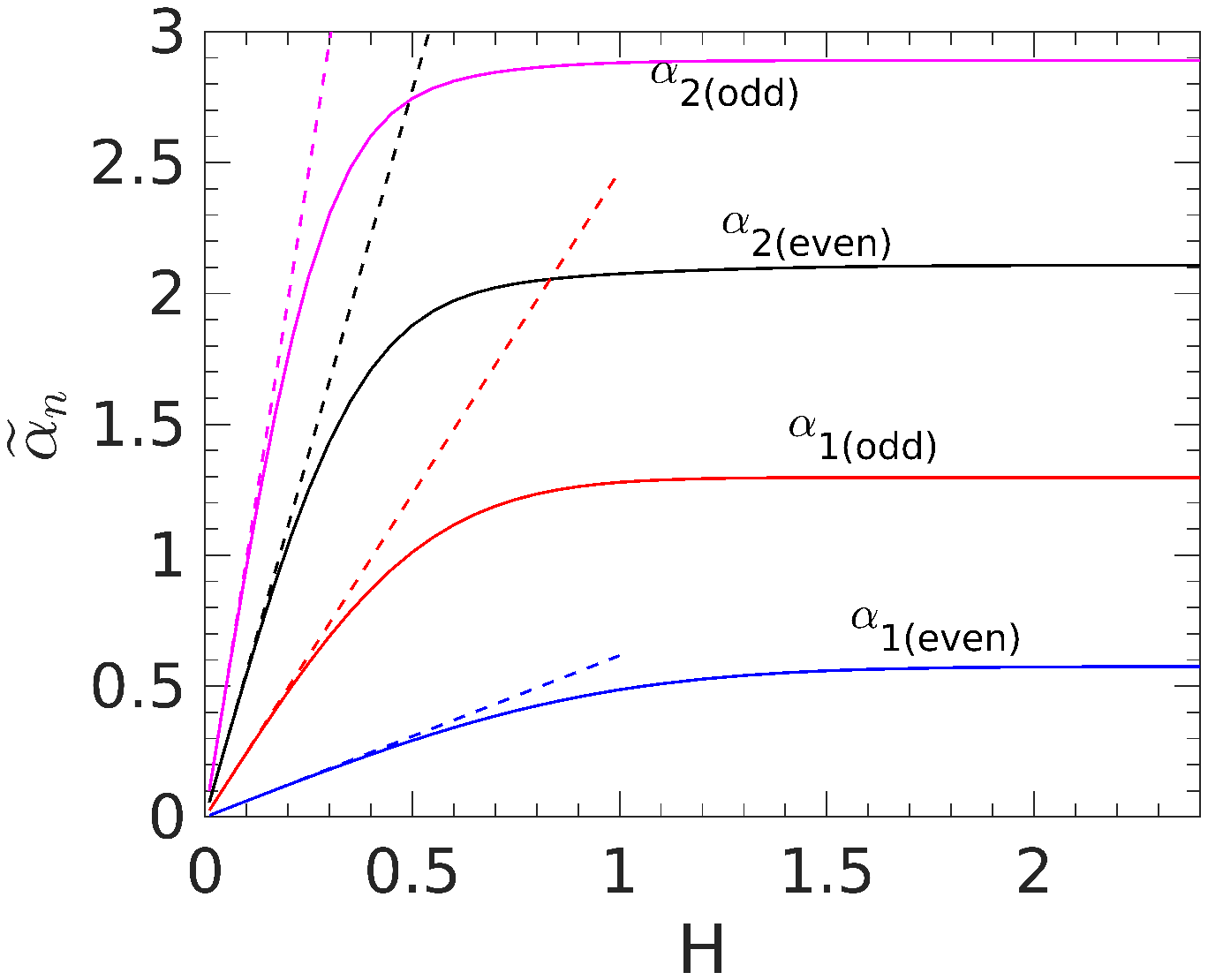}
		\caption{}
	\end{subfigure}
	\begin{subfigure}{0.495\textwidth}
		\includegraphics[trim={0cm 0cm 1cm 0cm},clip,width=\textwidth]{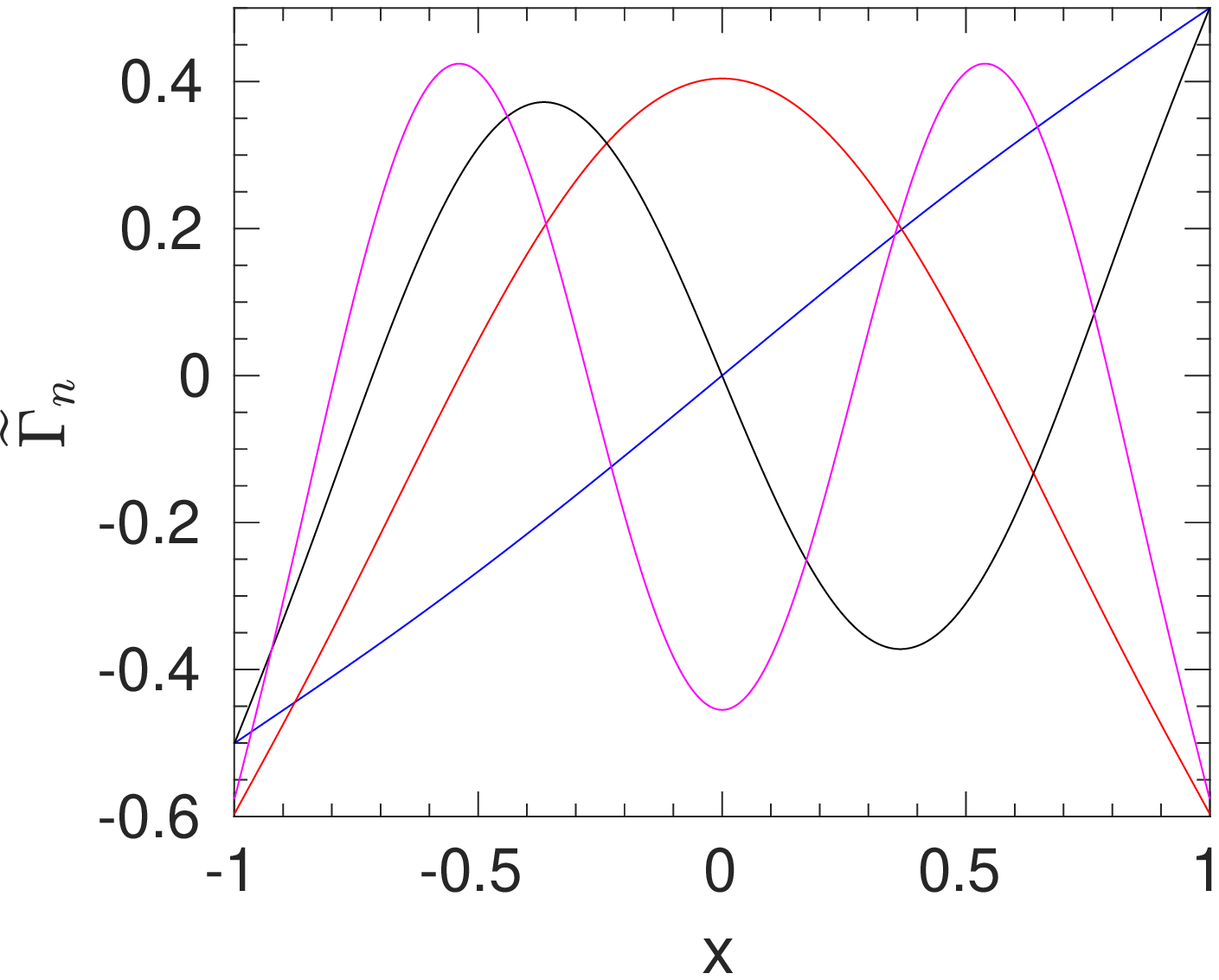}
		\caption{}
	\end{subfigure}
	\caption{(\textit{a}) Decay rates computed numerically by solving \eqref{eq:fd} (solid lines) as functions of $H$, and  compared to the analytical predictions obtained from lubrication theory (\ref{eq:alphalubapprox}) in the limit $H\to 0$ (dashed lines).  (\textit{b})  Plots of the surfactant concentration profiles (using the same colour code as in (\textit{a})) for the first two even and two odd modes for $H=2$ using a solution for the streamfunction  $\widetilde{\psi}_n$ calculated numerically using $4000\times4000$ gridpoints. 
	}
	\label{fig:Alphas}
\end{figure}

Figure~\ref{fig:Alphas}(\textit{a}) shows how the decay rates  $\widetilde{\alpha}_n$ of the odd and even modes, computed numerically from the eigenvalue problem (\ref{eq:fd}), vary with the depth of the domain $H$.  The decay rates become independent of the cavity depth for $H\gtrapprox 2$.  
We can find an exact asymptotic expression for the odd and even decay rates in the limit $H\to 0$ using a lubrication approximation  (see Appendix~\ref{app:thin})
\begin{equation}
    \alpha_n \to \frac{n^2\pi^2 H}{4}, \ \text{(odd modes)}, \quad \alpha_n \to \frac{(2n-1)^2\pi^2H}{16} \ \text{(even modes)},
    \label{eq:alphalubapprox}
\end{equation}
as shown with dashed lines in figure~\ref{fig:Alphas}(\textit{a}) for the first two odd and even modes. The surfactant concentration profiles of the corresponding modes are shown in figure~\ref{fig:Alphas}(\textit{b}) (using the same colours as  in figure~\ref{fig:Alphas}\textit{a}). All the surfactant profiles have a non-zero finite value and slope at the boundaries $x=\pm1$, as anticipated from (\ref{eq:Gamma}). We compute the dominant singularity strength $-4A_{n2}$  from the slope of the surfactant profile at the boundaries $x=\pm1$ according to (\ref{eq:Gamma}).

\begin{figure}
	\centering
	\begin{subfigure}{0.49\textwidth}
	        \begin{subfigure}{\textwidth}
	            \includegraphics[trim={0cm 0cm 1cm 0cm},clip,width=\textwidth]{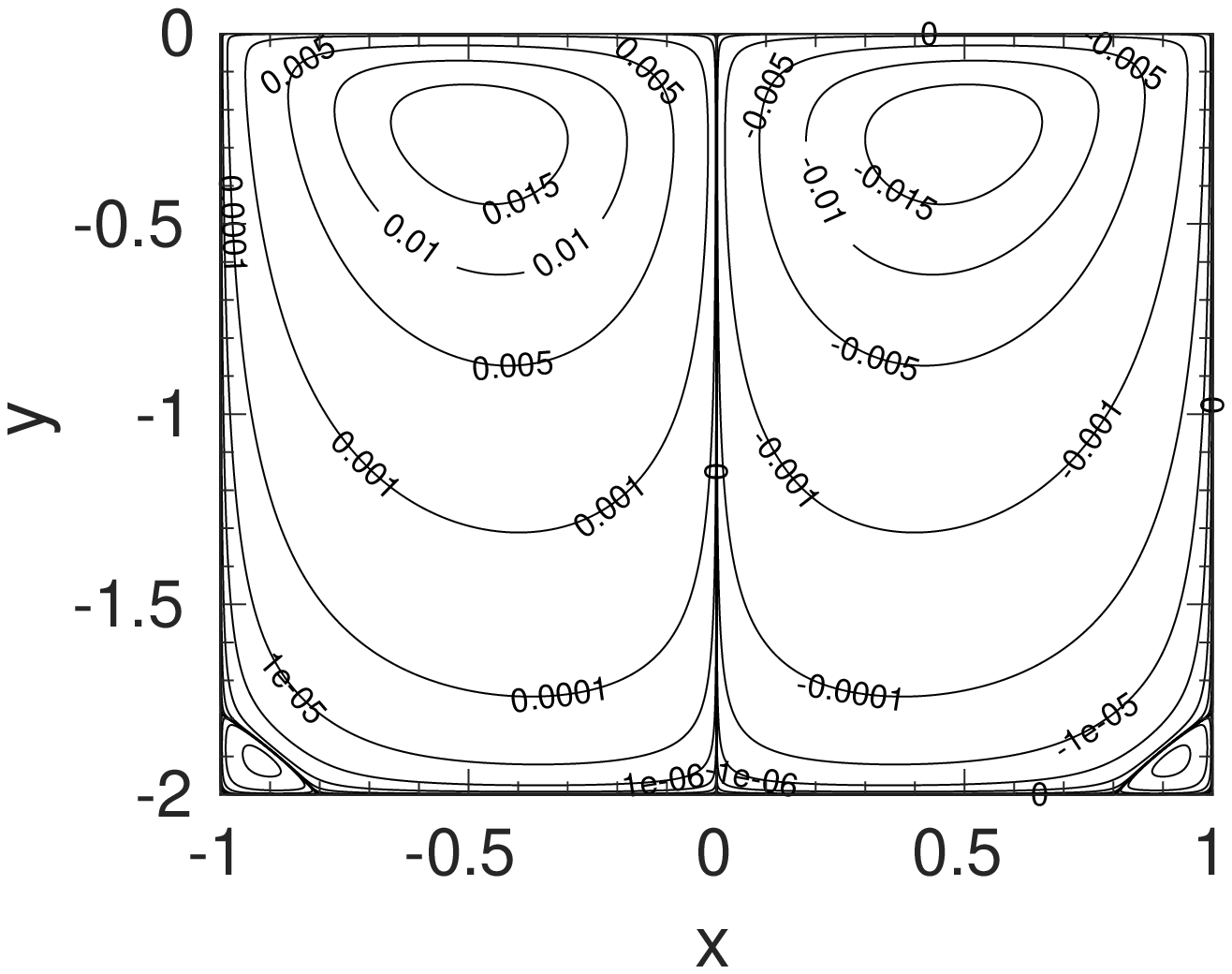}
		\vspace*{-5mm}
	       	\caption{}
	        \end{subfigure}
	        \begin{subfigure}{\textwidth}
	            \includegraphics[trim={0cm 0cm 1cm 0cm},clip,width=\textwidth]{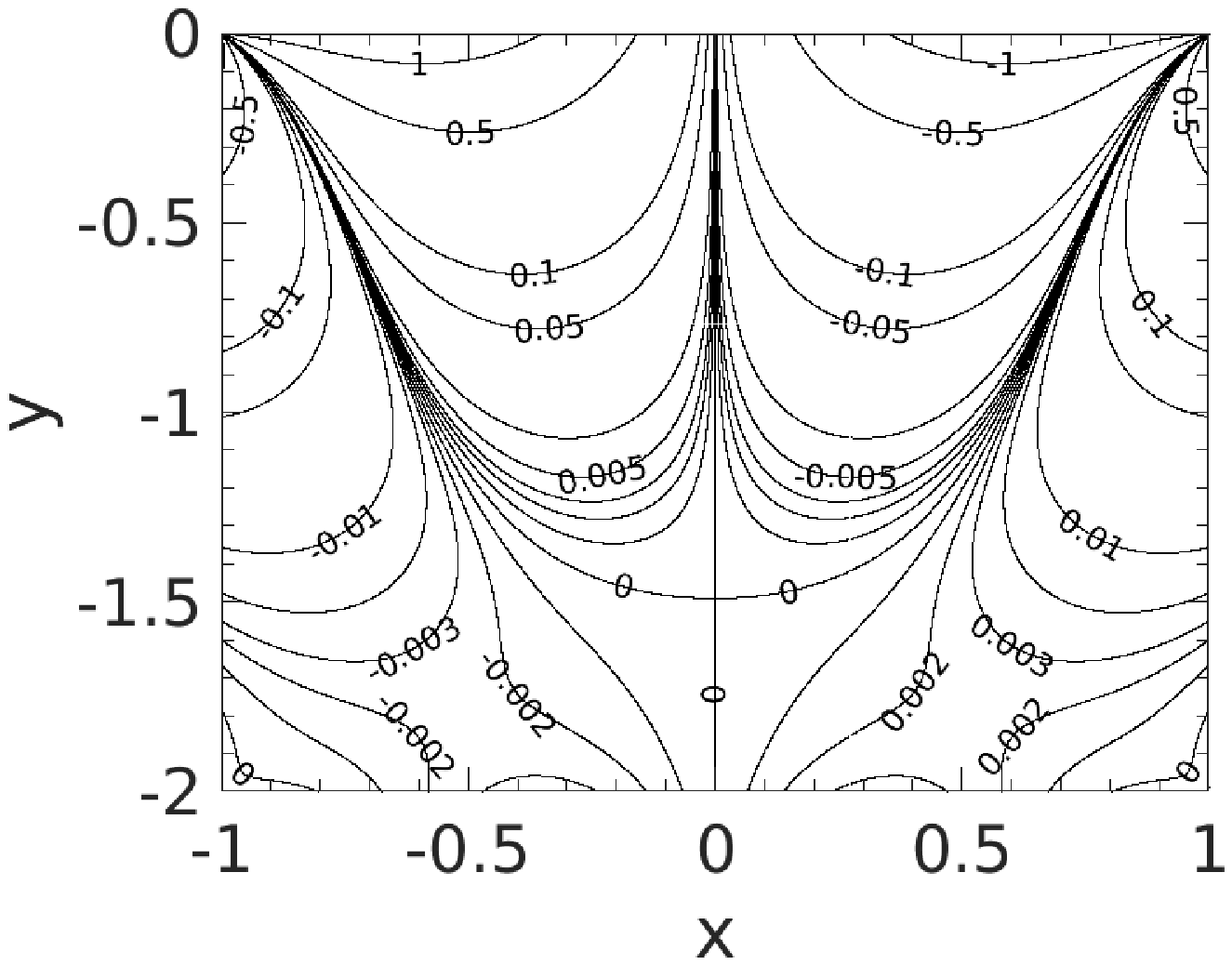}
		\vspace*{-5mm}
	       	\caption{}
	        \end{subfigure}
	\end{subfigure}
	\begin{subfigure}{0.37\textwidth}
    	\includegraphics[trim={0cm 0cm 1cm 0cm},clip,width=\textwidth]{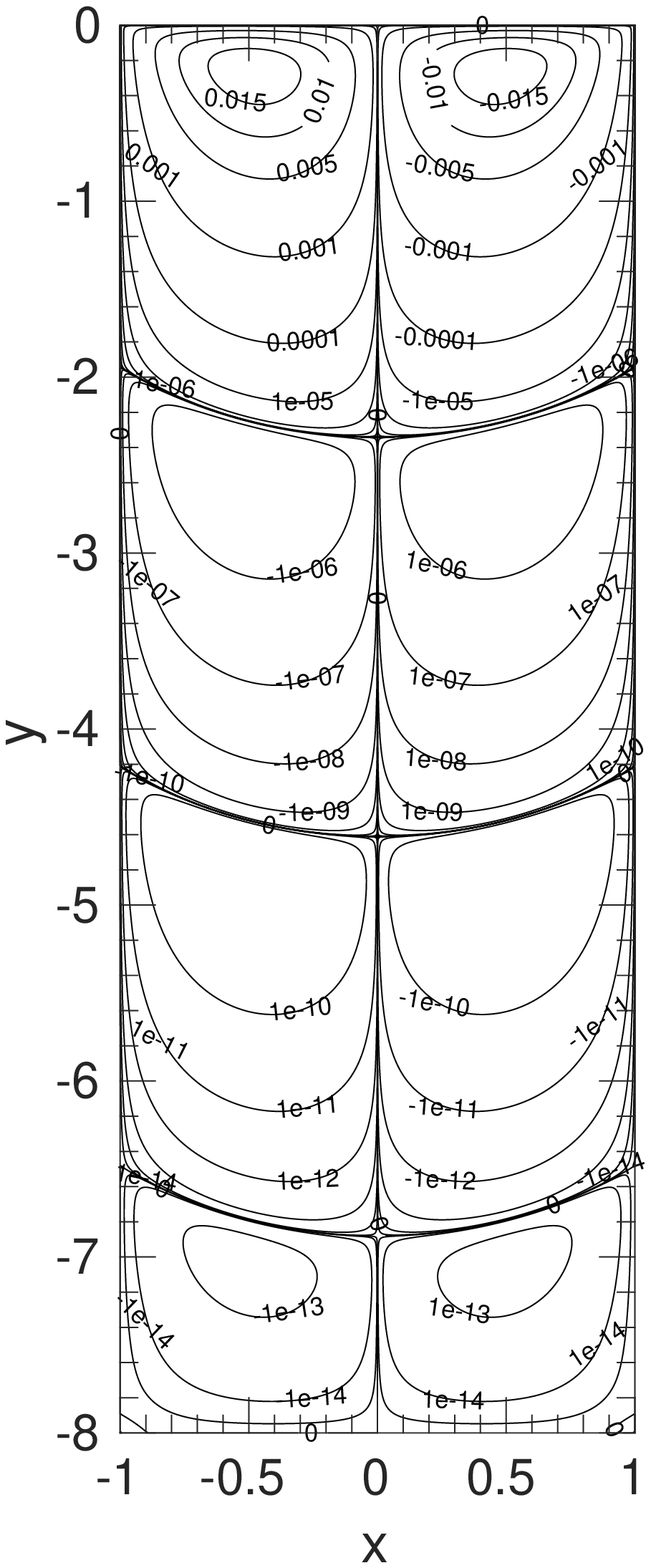}
    	\vspace*{-8mm}
    	\caption{}
	\end{subfigure}
	\begin{subfigure}{\textwidth}
	\centering
    	\includegraphics[trim={0cm 0cm 1cm 0cm},clip,width=\textwidth]{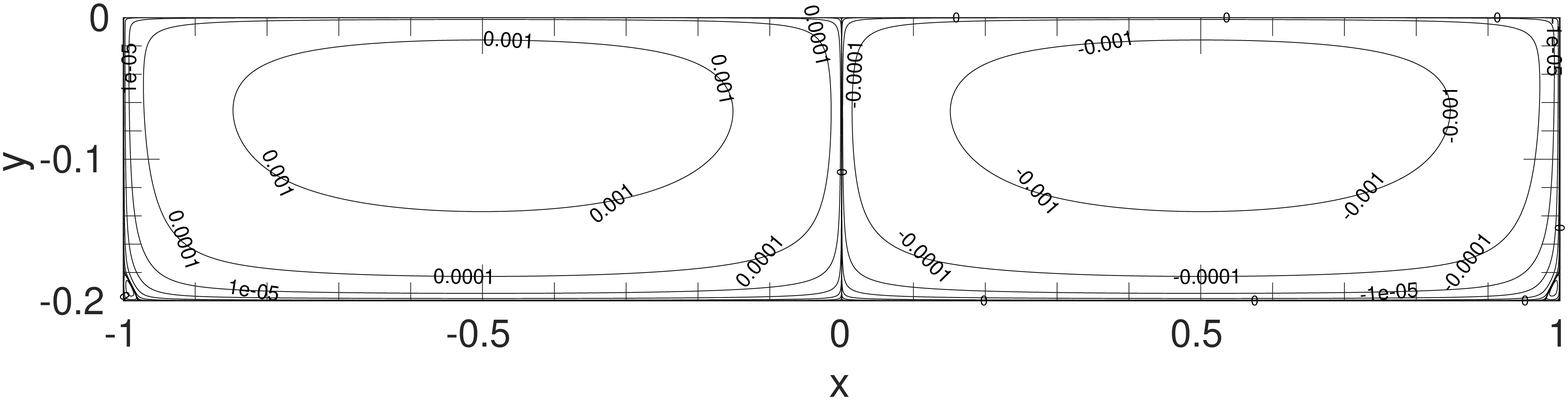}
    	\caption{}
	\end{subfigure}
	\caption{Contour plots of the dominant mode, the first  odd eigenmode ($n=1$), computed from numerical simulations in a square domain $(H=2)$, showing  (\textit{a}) the stream function and (\textit{b}) the vorticity. Similarly, (\textit{c}) shows numerical results of the contour plots of the stream function for the dominant mode (odd mode $n=1$) in a deep domain $(H=8)$ and (\textit{d}) a shallow domain $(H=0.2)$.  }
	\label{fig:firstmode}
\end{figure}

The contour plots computed numerically in figure~\ref{fig:firstmode}(\textit{a,b}) show the stream function and vorticity of the dominant mode (first odd eigenmode $n=1$) for $H=2$.  Weak Moffatt eddies can be seen in the lower corners of the cavity $(x,y)=(\pm 1,-2)$.  Vorticity contours converge at the upper corners, indicating that $\hat{\omega}$ is multivalued there, consistent with (\ref{eq:vortcorner}). The numerical results of the contour plot of the stream function in a deep channel with $H=8$, as shown in figure~\ref{fig:firstmode}(\textit{c}), reveals a sequence of recirculations. The strength of the stream function decreases rapidly with increasing depth, by approximately three orders of magnitude for the amplitude of the stream function  between successive cores.   In a shallow channel with $H=0.2$ (figure~\ref{fig:firstmode}\textit{d}), elongated eddies appear, consistent with predictions of lubrication theory.

\begin{figure}
	\begin{subfigure}{0.49\textwidth}
		\includegraphics[trim={0cm 0cm 1.75cm 0cm},clip,width=\textwidth]{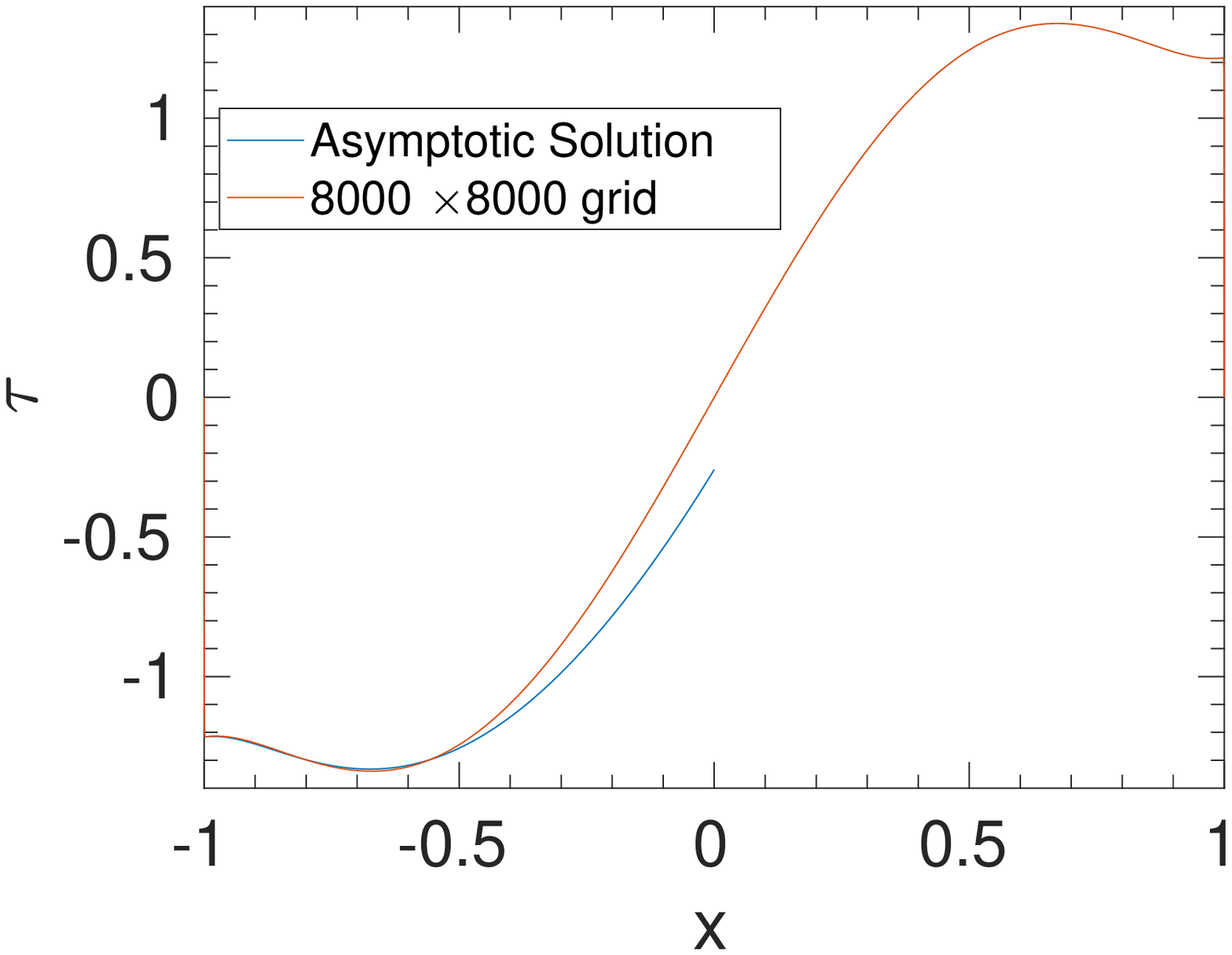}
		\caption{}
	\end{subfigure}
	\begin{subfigure}{0.49\textwidth}
		\includegraphics[trim={0cm 0cm 1cm 0cm},clip,width=\textwidth]{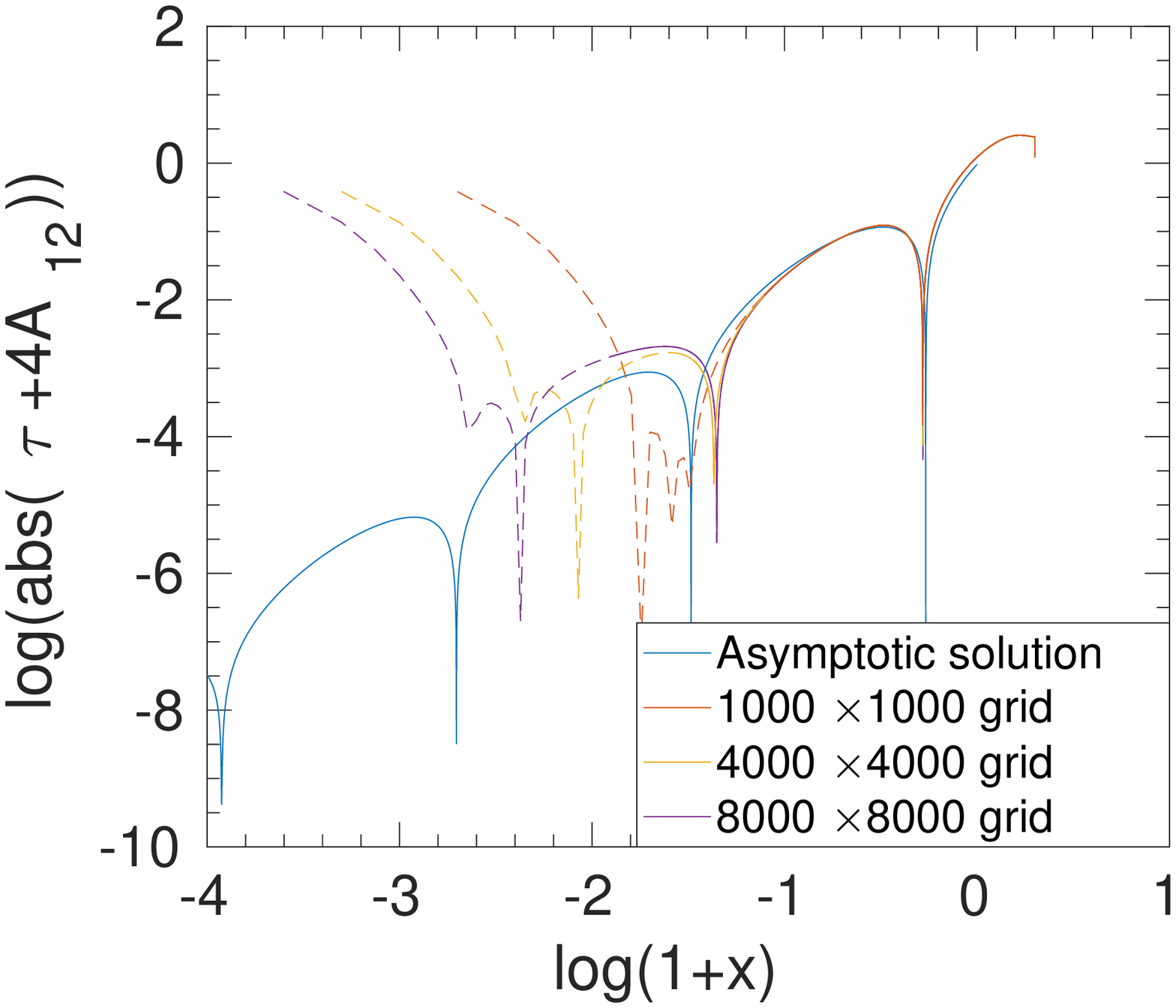}
		\caption{}
	\end{subfigure}
	\begin{subfigure}{0.49\textwidth}
		\includegraphics[trim={0cm 0cm 1cm 0cm},clip,width=\textwidth]{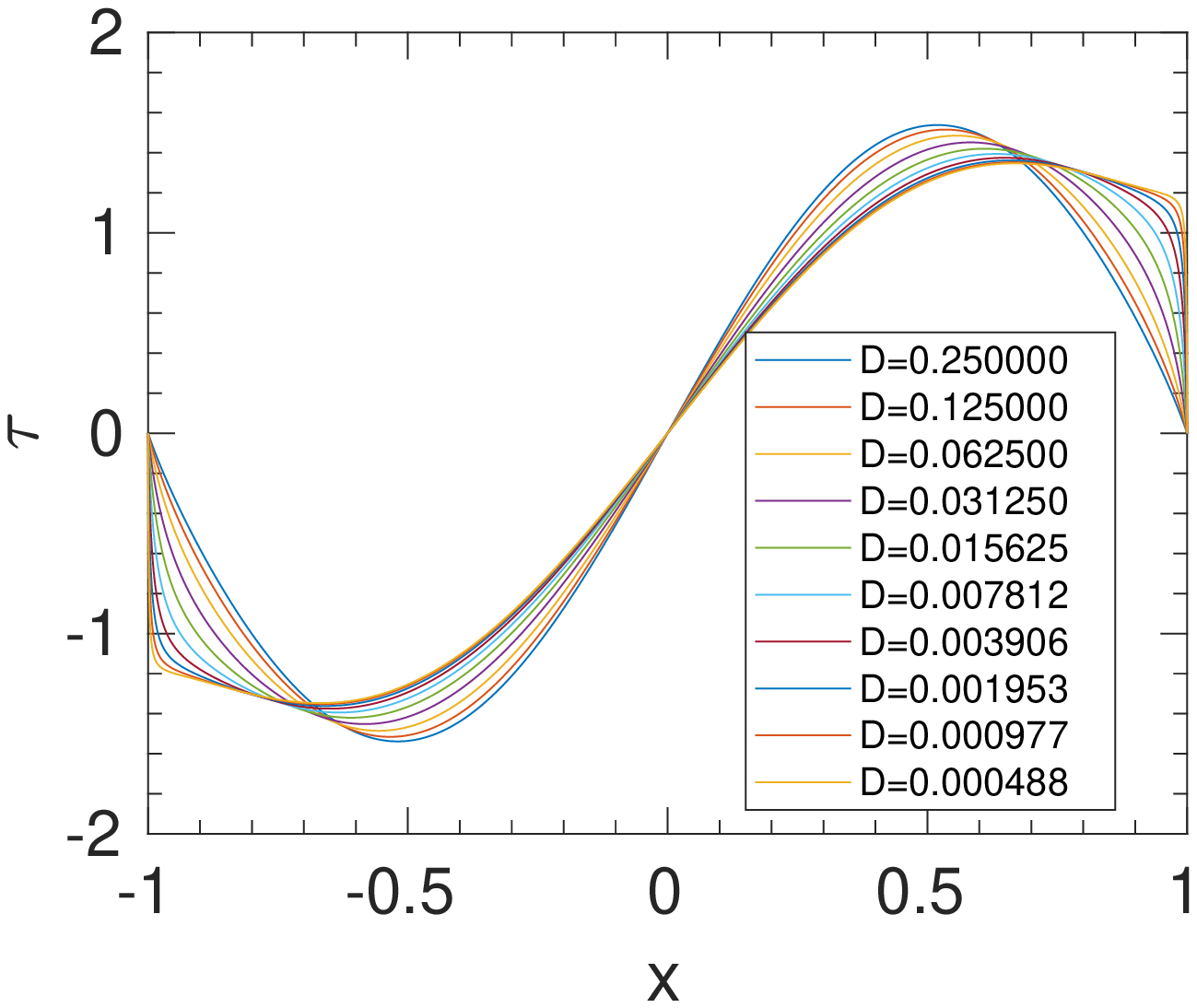}
		\caption{}
	\end{subfigure}
	\begin{subfigure}{0.49\textwidth}
		\includegraphics[trim={0cm 0cm 1cm 0cm},clip,width=\textwidth]{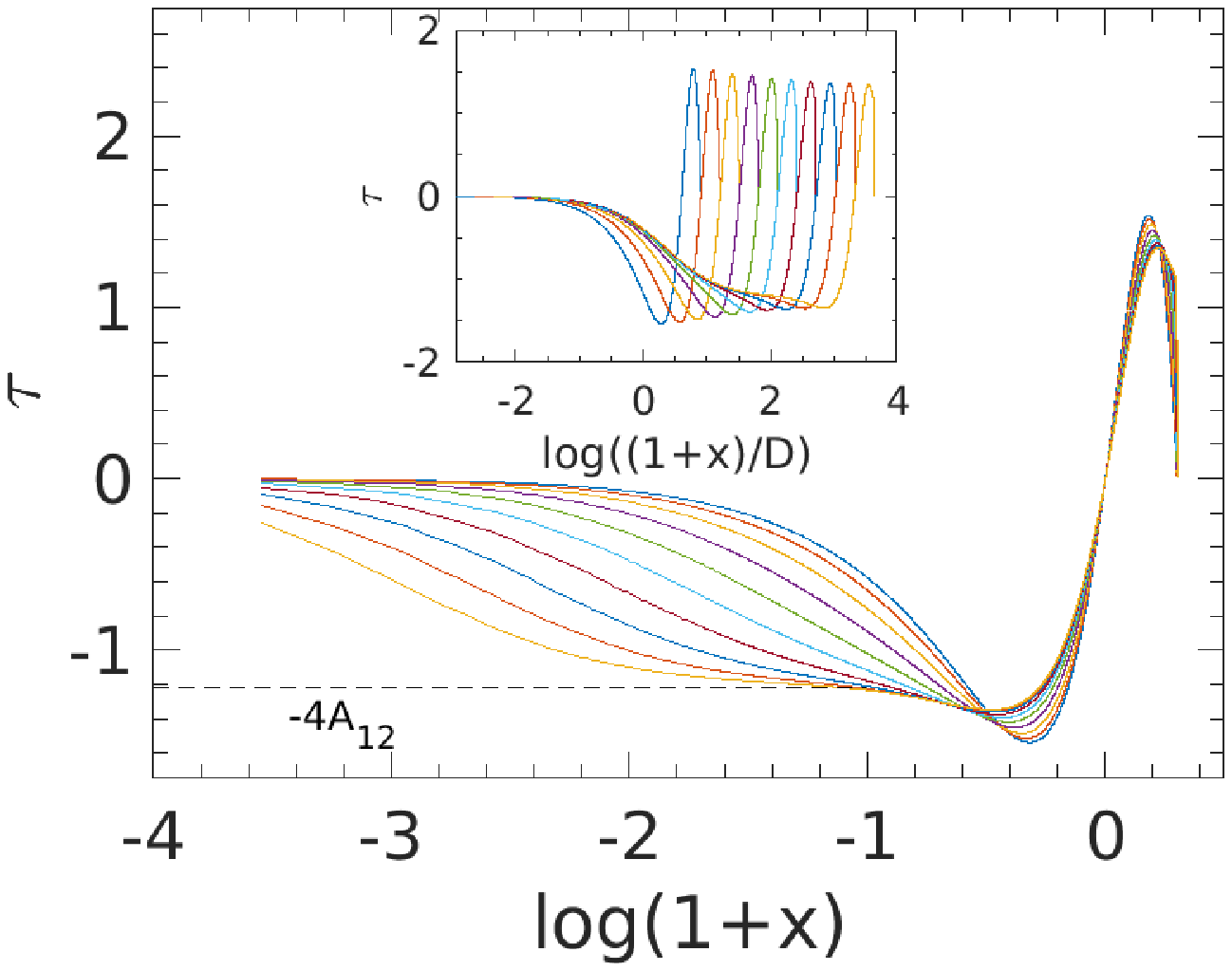}
		\caption{}
	\end{subfigure}
	\caption{Distribution of the surface shear stress computed from the dominant mode, the first odd mode $n=1$.  (\textit{a}) Numerical results (red) versus asymptotic results including two terms in (\ref{eq:shearwall}) (blue) with coefficients $4A_{12}=1.216$ and $A_{1a_{1}}\approx-0.175e^{0.157i}$ which were locally optimised to fit the numerical solution.  (\textit{b})  Logarithmic plot showing similar results as in (\textit{a}) versus $\log(1+x)$. This reveals overlap between asymptotics and numerics for different grid resolutions; dashed lines for the numerical results are used for $1+x\leq100\Delta x$, where numerical errors increase. (\textit{c}) Distributions of the shear stress, calculated numerically and when surface diffusion is included following \eqref{eq:diffusionBC}. (\textit{d}) Semi-log plots of the profiles in (\textit{c}), with the horizontal coordinate scaled by $D$ in the inset. The dashed line in the main graph is the asymptotic value of the shear stress at $x=-1$, i.e. $\tau=-4A_{12}=-1.216$}
	\label{fig:StressAsymp}
\end{figure}

Figure~\ref{fig:StressAsymp}(\textit{a}) shows the surface shear stress, computed numerically from the viscous-Marangoni stress condition $\widetilde{\tau}(x)=-\widetilde{\Gamma}_x(x)$, for the dominant mode (first odd mode, $n=1$) for $H=2$, revealing an oscillatory structure near the contact lines as $x\rightarrow\pm 1$, consistent with (\ref{eq:shearwall}).  The log--log plot in figure~\ref{fig:StressAsymp}(\textit{b}) reveals in more detail how the stress calculated from the numerical solution matches against the stress found using the asymptotic approximation (\ref{eq:shearwall}).  The finite difference approximation, with a finite grid size, necessarily fails to capture the increasingly short-wavelength oscillations as $x\rightarrow -1$ (as plotted with dashed lines when $1+x\leq100\Delta x$), and the asymptotic solution can be expected to fail as $1+x$ becomes too large.  However, there is an overlap region (indicated by solid lines for the numerical results), which grows in size with increasing grid resolution, over which the agreement is sufficiently strong to provide confidence in the numerical predictions throughout the rest of the domain. Thus, at the maximum grid resolution ($\Delta x=1/8000$) the numerical results are close to the asymptotic results for $\log(1+x)\approx -1.8$. We note that the asymptotic results could be made more accurate by including more terms in the series (\ref{eq:shearwall}), which would for instance show variations in the wavelength of the shear stress oscillations as $x\to-1$. However, the dominant odd mode $n=1$ clearly captures the oscillatory behaviour of the shear stress near the corner.

\subsection{Regularising the corner singularities}

The corner singularities can be regularised by adding a small amount of {surface} diffusion in the problem. In this case the transport equation for the surfactant, the first equation in (\ref{UnIntegratedBoundaryCond}\textit{a}),  modifies to
$\Gamma^*_{t^*} = -(\Gamma^* u^*)_{x^*} + D^*\Gamma^*_{x^*x^*}$,
in dimensional form with $D^*$ the surface diffusivity of the surfactant.
Hence, the stress boundary condition in the eigenvalue problem for the perturbation stream function (first equation in \eqref{eq:bc3})  becomes
\begin{equation}\label{eq:diffusionBC}
  \alpha\hat{\psi}_{yy} = -\hat{\psi}_{xxy} +D\hat{\psi}_{xxyy},
\end{equation}
where $D = D^*\mu^*/(W^*S^*\bar{\Gamma}$). We then specify an additional boundary condition and impose no-flux of surfactant, $\Gamma_x=0$, at the contact lines $(x,y)=(\pm1,0)$.  Thus, in the presence of weak surface diffusion, surface stress falls abruptly to zero in small boundary layers at the wall for ${D}\ll 1$ (figure~\ref{fig:StressAsymp}\textit{c}). Increasing ${D}$ causes the surface stress of the first odd mode $n=1$ to take a smoother more sinusoidal profile, revealing the impact of surface diffusion on the free surface.  The profile of the shear stress distribution near the contact line is shown in greater detail in figure~\ref{fig:StressAsymp}(\textit{d}) (using a logarithmic spatial scale), demonstrating its adjustment from the constant value $-4A_{n2}$ (value of the shear stress at the corner for $D=0$, plotted with a dashed line) to zero over a very short length scale.  Collapse of this data when $x$ is rescaled by ${D}$ (inset) provides evidence that weak surface diffusion regularises the singularity over a boundary layer of characteristic length scale ${D}$.

\begin{figure}
    \centering
    \begin{subfigure}{0.5\textwidth}
    \centering
   \includegraphics[trim={0cm 0cm 1cm 0cm},clip,width=\textwidth]{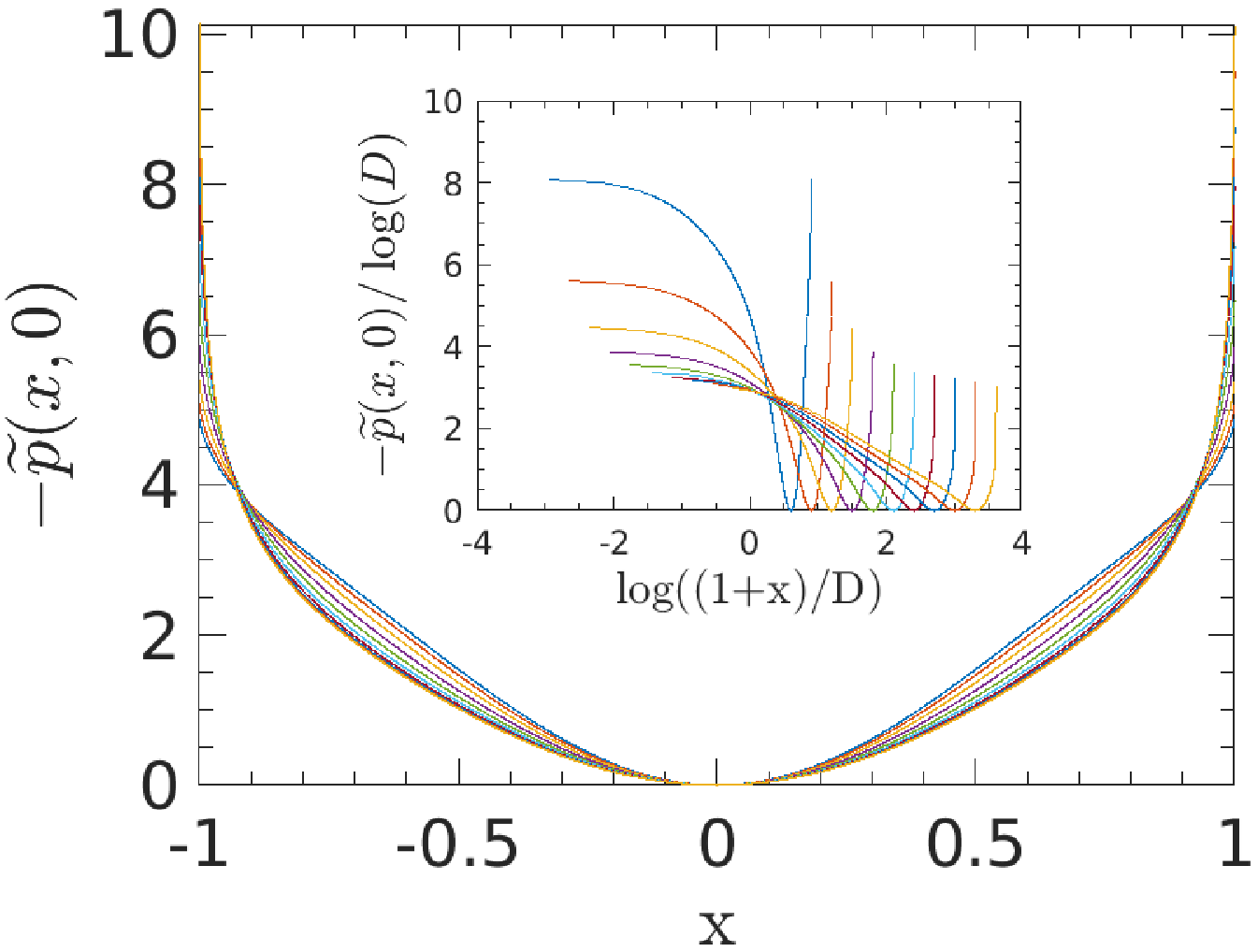}
    \caption{}
   \end{subfigure}
    \begin{subfigure}{\textwidth}
		\includegraphics[trim={0cm 0cm 1cm 0cm},clip,width=\textwidth]{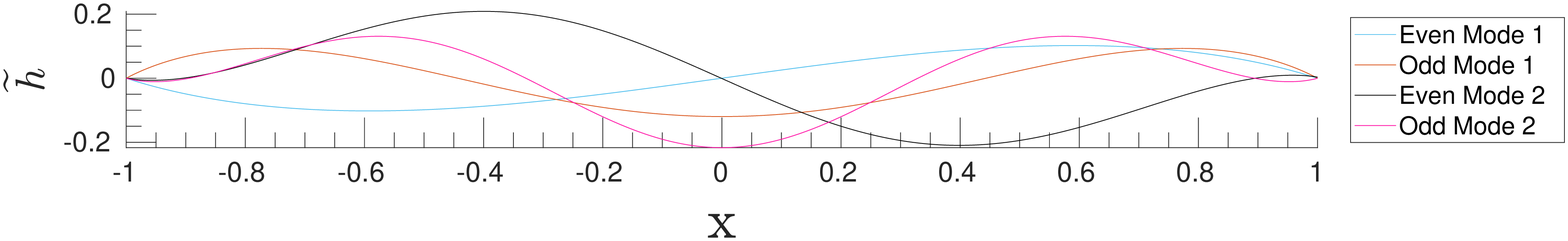}
		 \caption{}
	\end{subfigure}
	 \caption{(\textit{a}) Surface pressure profiles computed numerically for the different values of the diffusion constant $D$ given using the colours indicated in figure \ref{fig:StressAsymp}(\textit{c}). The inset shows a scaled semi-log graph showing the pressure profiles collapsing on to each other in a diffusive boundary layer located for  $x\in [-1,0]$ as $D$ decreases. (\textit{b}) Non-dimensional interfacial deflections (relative to the length-scale $W^* S^*/\gamma_0^*$), computed numerically as the leading-order correction to the flat state for the first two odd and even modes.}
	 \label{fig:pressure}
\end{figure}

We assumed initially that a restoring force is present that is sufficiently strong to suppress interfacial deflections by imposing a flat interface and ignoring the normal stress condition.  Given the singularity in the pressure at the contact lines (\ref{eq:presswall}), it is prudent to revisit this boundary condition.  Linearizing the normal stress condition around $y=0$, we can state this as $\widetilde{p}(x,0)-p_{ext}-2\widetilde{\psi}_{xy}(x,0)=\widetilde{h}_{xx}$, where the dimensional deflection is $(W^* S^*/\gamma_0^*)\widetilde{h}$ and $p_{ext}$ is a reference pressure (recall that, for the small surfactant concentration variations considered here, we can assume $S^*\ll \gamma_0^*$).  The displacement is computed by imposing $\widetilde{h}=0$ at each contact line, and imposing a volume constraint $\int_{-1}^1 \widetilde{h}\,\mathrm{d}x=0$. 
The pressure field at the free surface (computed numerically with gauge $\widetilde{p}(0,0)=0$) is shown in figure~\ref{fig:pressure}(\textit{a}) for the same values of D as used in figure \ref{fig:StressAsymp}(\textit{c,d}).  Collapse of the data (inset in figure~\ref{fig:pressure}\textit{a}) demonstrates that the pressure, like the stress, is regularised over a length scale in $x$ of $O(D)$ at the contact lines, reaching a maximum value of $O(\log(1/D))$.  The corresponding interfacial displacement (figure~\ref{fig:pressure}\textit{b}), computed numerically with $D=0$, shows that, despite the strong local forcing, displacements remain bounded.  For the first odd mode, for example, there is weak upwelling near each contact line compensated by lowering of the free surface in the middle of the domain.

\section{Discussion}
\label{sec:disc}

Confined gas--liquid interfaces are commonly contaminated by surfactant, deliberately or by trace amounts naturally present in the environment.  Here, we have addressed Marangoni-driven spreading of insoluble surfactant, towards an equilibrium state with uniform concentration, taking place across the width of an interface in a channel, when viscosity dominates the flow.  Many features of this spreading are diffusive in character, particularly the decomposition of the flow into a set of mutually orthogonal eigenmodes.  The modes decay exponentially in time at different rates, with the longest-wave modes being the most long-lasting.  Despite this benign temporal structure, the dynamic compression of surfactant near each lateral boundary gives eigenmodes more exotic spatial features.  The logarithmic pressure singularity near each contact line (\ref{eq:presswall}) has the potential to generate a measurable surface deflection (figure~\ref{fig:pressure}), while the shear stress has a pronounced oscillatory structure (figure~\ref{fig:StressAsymp}\textit{b}).

So far, we have focussed our study to the special case of a single-fluid flow with a pinned contact line forming a $\pi/2$ contact angle with the cavity walls. However, we can relax these assumptions to show that our results extend to a broader class of problems. In Appendix \ref{app:Exponents}, we present a local asymptotic analysis for the case of a two-fluid incompressible Stokes flow, with fluids of arbitrary viscosities and with an arbitrary contact angle between 0 and $\pi$ for the interface which is still assumed locally flat and pinned to the flat walls of the cavity. We show that the structure of the (two-fluid) flow near the contact line and the type of singularities generated by the time-dependent spreading of the surfactants lying at the interface between the two fluids is similar to what we  found for the single-fluid flow with $\pi/2$ contact angle. Indeed, the  singularity is always associated with a real exponent of $2$ in the $r$-dependence of the streamfunction, which leads to the logarithmic pressure singularity and the multi-valuedness of the vorticity  near each contact line. The main difference is that the part of the series of the streamfunction associated with real exponents does not terminate at $r^3$ for contact angles different from $\pi/2$ (see \eqref{eq:realseriesPsi} for the streamfunction with $\pi/2$ contact angle). The oscillatory structure of the shear stress, associated with complex exponents in the $r$-dependence of the streamfunction,  depends on the wedge angle, but not on the viscosity ratio between the two fluids. The viscosity ratio only affects the coefficients of higher order terms in the series for each eigenmode. We note that a local analysis is sufficient to establish the local flow structure and type of singularities near the contact line, which can be generated by an arbitrary far-field disturbance of the surfactant distribution. Such fundamental surfactant flows are found across the range of applications mentioned in the introduction.

We have also shown how, in the case of a single-fluid flow with $\pi/2$ contact angle, the introduction of a small amount of surface diffusion regularises the contact line singularities, leading to prominent changes in stress distributions (figure~\ref{fig:StressAsymp}c,d).  Surface diffusivity of $7\times 10^{-10}\,$m$^2$/s for the common surfactant sodium dodecylsulfate (SDS) \citep{CHANG19951} translates to a dimensionless diffusion coefficient $D=\mu^* D^*/(S^* W^*)$ below $10^{-8}$ in magnitude, assuming a spreading coefficient $S^*=10^{-2}\,$kg/s$^2$ over water in a channel of width 1 cm.  At such low levels, the impact of the singularities may still be visible, with the pressure maximum near each contact line, proportional to $\log(1/D)$, generating surface deflections resembling that shown in figure~\ref{fig:pressure}. We expect that diffusion will act in the same way for the broader class of problems involving two-fluid viscous flows and arbitrary contact angles between 0 and $\pi$.

The singular flow structures near contact lines that we have identified have the potential to contaminate dynamic computations that do not take proper account of these small-scale local structures.  In our calculations of spatial eigenmodes, we chose to combine asymptotic analysis with a dense numerical grid to capture the dominant spatial features of the flow.  There are a range of alternative strategies that could be deployed, notably singularity removal \citep{sprittles2011,game2019}, although (at least) two distinct singularities would require removal in the present problem.  As indicated above, singularity regularisation via the introduction of surface diffusion can operate over extremely small lengthscales when using realistic parameter values, itself presenting a computational challenge.  Singularities can be expected to become a particular difficulty in time-dependent studies, when artificial diffusion associated with a computational scheme may generate spurious disturbances propagating outward from the contact lines.  Corner singularities can also present convergence difficulties for numerical schemes that represent solutions using (Fadle--Papkovich) eigenfunctions that assume separability of spatial variables \citep{meleshko96, meleshko1997}.

The present model rests on numerous assumptions.  We have restricted attention to the near-equilibrium state, ignoring nonlinearities associated with large concentration gradients.  One benefit of this assumption is that small concentration changes support our assumption of a linearized equation of state, and the assumption that the Marangoni flow is sufficiently weak for restoring forces to maintain a nearly flat interface.  We have assumed that the surfactant is insoluble, but anticipate that desorption near the contact line may contribute to regularisation of the singularity. While the planar flow studied here could readily be extended to an axisymmetric geometry, a potentially more interesting avenue, with regard to three-dimensionality, will be to examine how the transverse flow studied here interacts with axial flows along the channel, as may occur in plastrons used in superhydrophic drag reduction \citep{peaudecerf2017}, in maze solving by surfactant \citep{temprano2018soap}, or in microfluidic applications. Furthermore, although we have considered a purely viscous regime, Marangoni spreading can be very rapid.  The dimensional decay rates predicted here are $1/\alpha$ times $W^* \mu^*/ \overline{\Delta \gamma}^*$, where $\alpha$ is an $O(1)$ modal decay rate and  $\overline{\Delta\gamma}^*=(\gamma_0^*-\gamma_c^*)\bar{\Gamma}^*/\Gamma_c^*$ is the surface tension reduction due to the equilibrium surfactant distribution.  Taking this as low as $10^{-3}$kg/s$^2$, say, for water in a narrow channel of width 1 mm (and comparable depth), the decay timescale is approximately 1 ms$/\alpha$, with a Reynolds number of order unity. Despite decaying exponentially with respect to time, the structure of the flow in wider and deeper channels can therefore be expected to be influenced by inertia, at least initially.


In summary, this study shows how the unsteady spreading of a surfactant monolayer along a liquid-liquid or liquid-gas interface, confined by a lateral rigid boundary, can generate singular flow structures near stationary contact lines, including a logarithmically divergent pressure field and an oscillatory shear stress distribution.  Careful treatment of these structures is needed in computational simulations involving dynamic surfactant transport in confined domains.  


\section*{Acknowledgements}
The authors are grateful to Prof.~David Silvester for enlightening discussions relating to the numerical methods used in this study.

\vspace{0.5cm}
\textbf{Declaration of Interests.}
The authors report no conflict of interest.

\appendix
\section{The energy budget}
\label{app:energy}

The Stokes equation can be written as $\boldsymbol{\nabla} \cdot \boldsymbol{\sigma} = \mathbf{0}$, where the non-dimensional Cauchy stress tensor is $\boldsymbol{\sigma} = -p\mathbf{I} + 2 \mathsf{e}$ and the strain-rate tensor $\mathsf{e} = (\boldsymbol{\nabla}\mathbf{u} + \boldsymbol{\nabla}\mathbf{u}^T)/2$. Thus, for a Stokes flow, 
$\boldsymbol{\nabla}\cdot (\mathbf{u}\cdot \boldsymbol{\sigma}) 
=  \boldsymbol{\sigma}: \boldsymbol{\nabla } \mathbf{u} = \boldsymbol{\sigma}:\mathsf{e}$, exploiting the symmetry of $\boldsymbol{\sigma}$.  Given 
that $p\mathbf{I}:\mathbf{e} = \text{trace}(\mathsf{e})p = 0$ by incompressibility, it follows that
	$\boldsymbol{\nabla}\cdot (\mathbf{u}\cdot \boldsymbol{\sigma}) = 2\mathsf{e}: \mathsf{e}$. 
Integrating this over the domain and applying the divergence theorem gives
\begin{equation}\label{eq:viscousworkbalance}
	2\int_V \mathsf{e :e} \ \mathrm{d}A = \int_{\partial V} \mathbf{u} \cdot \boldsymbol{\sigma} \cdot {\mathbf{n}} \ \mathrm{d}s,
\end{equation}
where $\partial V$ represents the boundary of $V$ and $\mathrm{d}s$ is the curvilinear element along the boundary. For the present perturbation problem, described by equations (\ref{eq:bih}), (\ref{UnIntegratedBoundaryCond}\textit{b,c}) and (\ref{eq:bc3})  for each eigenmode, equation (\ref{eq:viscousworkbalance}) balancing work done by Marangoni forces with viscous dissipation becomes
%
\begin{equation}\label{eq:workbalance1}
2  \int_V \left(2\hat{\psi}_{yx}^2 + \frac{1}{2}\hat{\psi}_{yy}^2+\frac{1}{2}\hat{\psi}_{xx}^2  -\hat{\psi}_{xx}\hat{\psi}_{yy} \right) \ \mathrm{d}A  
= -\int_{-1}^{1} \left. \left(\hat{\Gamma}_x \hat{\psi}_y \right) \right|_{y=0} \ \mathrm{d}x .
\end{equation}   
Integration by parts of the left hand side of (\ref{eq:workbalance1}) and the use  of the no-slip boundary condition on vertical boundaries and no penetration boundary condition on horizontal boundaries gives
\begin{equation}
	2  \int_V \left(2\hat{\psi}_{yx}^2 + \frac{1}{2}\hat{\psi}_{yy}^2+\frac{1}{2}\hat{\psi}_{xx}^2  -\hat{\psi}_{xx}\hat{\psi}_{yy} \right) \ \mathrm{d}A =   \int_V \left(  \hat{\psi}_{yy}^2+\hat{\psi}_{xx}^2  +2\hat{\psi}_{xx}\hat{\psi}_{yy} \right) \ \mathrm{d}A  = \int_V \hat{\omega}^2 \ \mathrm{d}A,
\end{equation} 
where $\hat{\omega} =-(\hat{\psi}_{yy} + \hat{\psi}_{xx})$ is the vorticity.  Further integration by parts of the right-hand-side of \eqref{eq:workbalance1} gives 
\begin{equation}
	- \int_{-1}^{1} \left. \left(\hat{\Gamma}_x \hat{\psi}_y \right) \right|_{y=0}   \ \mathrm{d}x  = \int_{-1}^{1} \left. \left(\hat{\Gamma} \hat{\psi}_{xy}  \right) \right|_{y=0} \ \mathrm{d}x .
\end{equation} 
Since $\hat{\psi}_{xy} = \alpha \hat{\Gamma}$ from (\ref{eq:bc4}), we obtain (\ref{eq:energy})
for each decay rate $\alpha$ and corresponding eigenmode $\{\hat{\psi},\hat{\omega},\hat{\Gamma}\}$.



\section{Orthogonality of modes}
\label{app:ortho}

The reciprocal theorem \citep{masoud_stone_2019} states that two Stokes flows, with velocity fields and Cauchy stress tensors $(\mathbf{u},\boldsymbol{\sigma})$ and  $(\mathbf{u}',\boldsymbol{\sigma}')$,    satisfy in the same region
\begin{equation}
	\int_{\partial V} \mathbf{u} \cdot(\boldsymbol{\sigma}'\cdot \mathbf{n})\, \mathrm{d}A = 
	\int_{\partial V} \mathbf{u}' \cdot(\boldsymbol{\sigma}\cdot \mathbf{n})\, \mathrm{d}A
\label{eq:b1}
\end{equation} 
For the present perturbation problem, described by equations (\ref{eq:bih}), (\ref{UnIntegratedBoundaryCond}\textit{b,c}) and (\ref{eq:bc3}) for each eigenmode, we have $\mathbf{u}=\mathbf{0}$ on the three solid boundaries whilst at the free surface  $u_y = - \Gamma_x$. Thus, for two distinct modes $m$ and $n$, (\ref{eq:b1}) implies
\begin{equation}
	\int_{-1}^1 \hat{u}_m \hat{\Gamma}_{n,x}\, \dx = 	\int_{-1}^1 \hat{u}_n \hat{\Gamma}_{m,x}\, \dx.
\end{equation} 
Integrating by parts and using $\alpha \hat{\Gamma} = -\hat{u}_x$ gives
\begin{equation}
	\left[\hat{u}_m\hat{\Gamma}_n\right]_{-1}^1  - \int_{-1}^1 \alpha_m\hat{\Gamma}_m\hat{\Gamma}_n \dx = \left[\hat{u}_n\hat{\Gamma}_m\right]_{-1}^1  - \int_{-1}^1 \alpha_n\hat{\Gamma}_n\hat{\Gamma}_m \dx.
	\label{eq:b3}
\end{equation} 
As the surface velocity is zero at $x=\pm1$, (\ref{eq:b3}) becomes
\begin{equation}
	(\alpha_m-\alpha_n)\int_{-1}^1 \hat{\Gamma}_m \hat{\Gamma}_n\, \dx = 0,
\end{equation} 
which shows that the surfactant concentration profiles corresponding to different modes are orthogonal since $\alpha_m \neq \alpha_n$, as stated in (\ref{eq:orthcond}). Equivalently, we note that this result can be derived by exploiting the self-adjointness of (\ref{eq:bih}), (\ref{UnIntegratedBoundaryCond}\textit{b,c}) and (\ref{eq:bc3}).

Numerical computation of the scalar product of the surfactant concentration modes for $1\leq m,n\leq 5$ gives \small
	\begin{equation}
	\int_{-1}^1 \widetilde{\Gamma}_m \widetilde{\Gamma}_n \ \mathrm{d}x = \mathbf{A}_{m,n} = 
	  \begin{pmatrix}
		0.20243519  & -0.00000524 & -0.00000370 & -0.00000345 & -0.00000358\\
  -0.00000524 &  0.19832581 & -0.00000394 & -0.00000457 & -0.00000508\\
  -0.00000370 & -0.00000394 &  0.19136202 & -0.00000570 & -0.00000635\\
  -0.00000345 & -0.00000457 & -0.00000570 &  0.19112895 & -0.00000744\\
  -0.00000358 & -0.00000508 & -0.00000635 & -0.00000744 &  0.19096022
	\end{pmatrix} . 
	\label{eq:ortho}
	\end{equation} \normalsize
These results were computed using our numerical scheme presented in section \ref{subsec:finitedifference} with $4000\times4000$ gridpoints. The non-diagonal elements of the matrix $\mathbf{A}_{m,n}$ in (\ref{eq:ortho}) are very small, showing the orthogonality of the modes calculated numerically and the global accuracy of our numerical scheme.

\section{The thin-film limit}
\label{app:thin}

As $H\rightarrow 0$, the biharmonic equation \eqref{eq:bih} can be approximated as $\psi_{yyyy} = 0$, therefore the general solution in this limit is given by
\begin{equation}
\psi = f_1(x)y^3 + f_2(x)y^2+ f_3(x)y + f_4(x).
\end{equation} The boundary condition $\psi(x,y=0) = 0$ means $f_4(x)=0$. The boundary conditions $\psi(x,y = -H) = 0$ and $\psi_y(x,y=-H)=0$ give 
\begin{equation}
 0 = -f_1(x)H^3 + f_2(x)H^2 -f_3(x)H,
\end{equation} and
\begin{equation}
 0 = 3f_1(x)H^2 - 2
 f_2(x)H +f_3(x).
\end{equation} Eliminating $f_3$ 
gives
$f_2(x)= 2f_1(x)H$,
implying
$f_3(x) = f_1(x)H^2$.
Hence
$\psi = f_1(x)(y^3+2Hy^2+H^2y)$.
The stress boundary condition at the surface $y=0$ is $\alpha\psi_{yy} = - \psi_{xxy}$, which imposes
\begin{equation}
4\alpha H f_1(x) = - f_1^{\prime \prime}(x)H^2.
\end{equation} The solution of this ordinary differential equation for $f_1(x)$ is
\begin{equation}\label{eq:f1condition}
 f_1(x) = C_1 \cos{\left(\sqrt{\frac{4 \alpha}{H}}x\right)} + C_2 \sin{\left(\sqrt{\frac{4 \alpha}{H}}x\right)} ,
\end{equation} 
for some arbitrary integration constants $C_1$ and $C_2$. The odd and even modes are given by setting either of these constants to zero. Let $C_2=0$, then applying the boundary conditions $\psi(x = \pm1)=0$ at the side walls gives
$\cos{\left(\sqrt{{4 \alpha}/{H}}\right)} = 0$,
which implies
\begin{equation}\label{eq:alphaevenlub}
    \alpha_n = \frac{(2n-1)^2 \pi^2 H}{16},
\end{equation} 
where $n$ can be any positive integer, $n\geq 1$. Equation \eqref{eq:alphaevenlub} is an approximation for the decay rates for the  modes even of the stream function in $x$ as $H$ becomes small. Similarly, letting $C_1=0$ in \eqref{eq:f1condition}, and then applying the boundary conditions $\psi(x = \pm1)=0$ at the side walls gives
$\sin{\left(\sqrt{\frac{4 \alpha}{H}}\right)}  = 0$,
which implies
\begin{equation}\label{eq:alphaoddlub}
    \alpha_n = \frac{n^2\pi^2 H}{4},
\end{equation}
where $n$ can be any positive integer, $n\geq 1$. Equation \eqref{eq:alphaoddlub} is an approximation for the decay rates for the  modes even of the stream function in $x$ as $H$ becomes small. We note finally that the sinusoidal modes (\ref{eq:f1condition}) in this long-wave theory predict zero slope of the surfactant concentration at the contact lines, so fail to capture the finite slope predicted by (\ref{eq:Gamma}) and demonstrated in figure~\ref{fig:Alphas}(\textit{b}).

\section{Asymptotics with arbitrary contact angle and two fluids of arbitrary viscosities}
\label{app:Exponents}

\begin{figure}
    \centering
    \begin{subfigure}{0.40\textwidth}
    \centering
   \includegraphics[trim={0cm 0cm 1cm 0cm},clip,width=\textwidth]{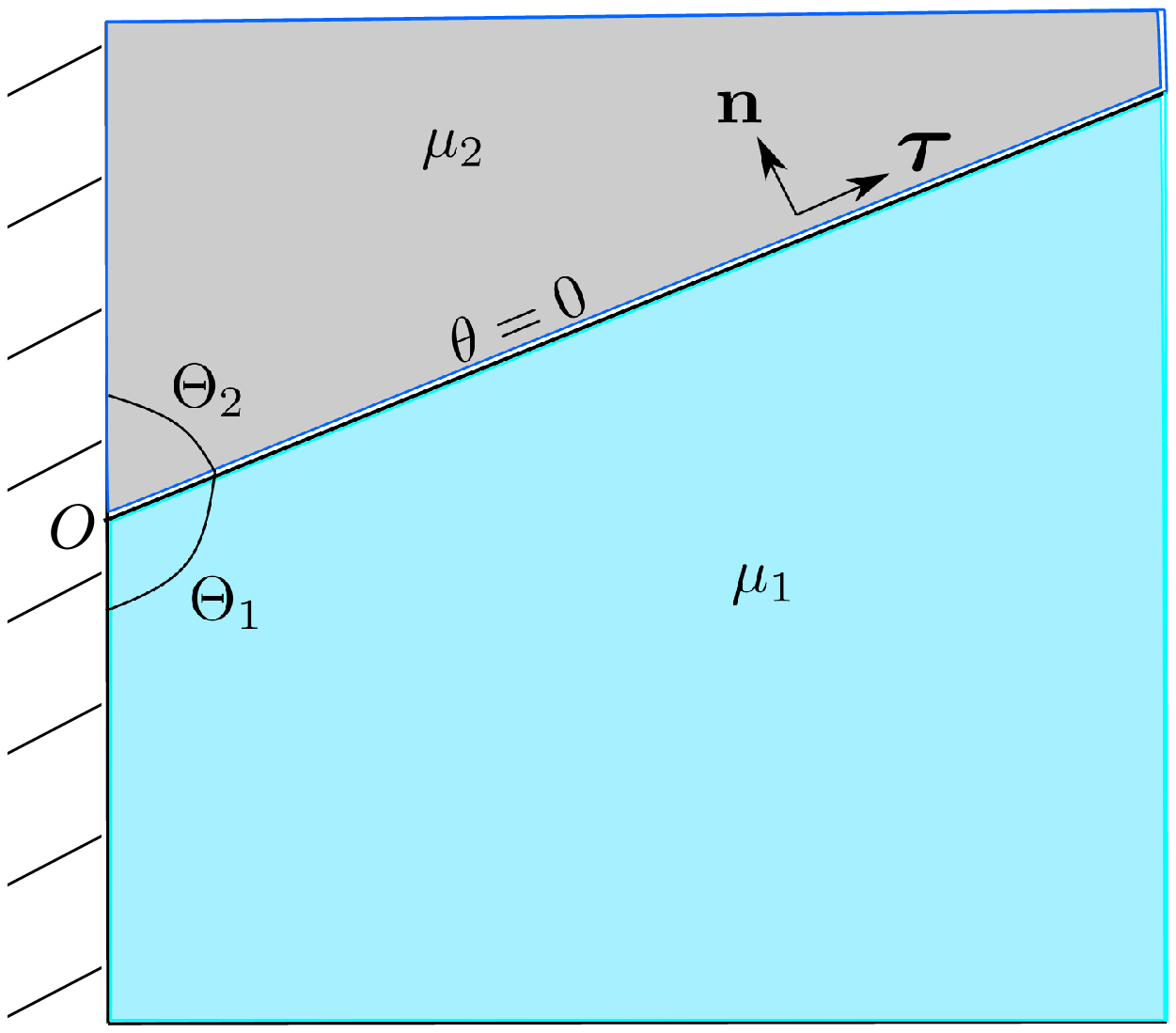}
    \caption{}
   \end{subfigure}
    \begin{subfigure}{0.52\textwidth}
		\includegraphics[trim={0cm 0cm 1cm 0cm},clip,width=\textwidth]{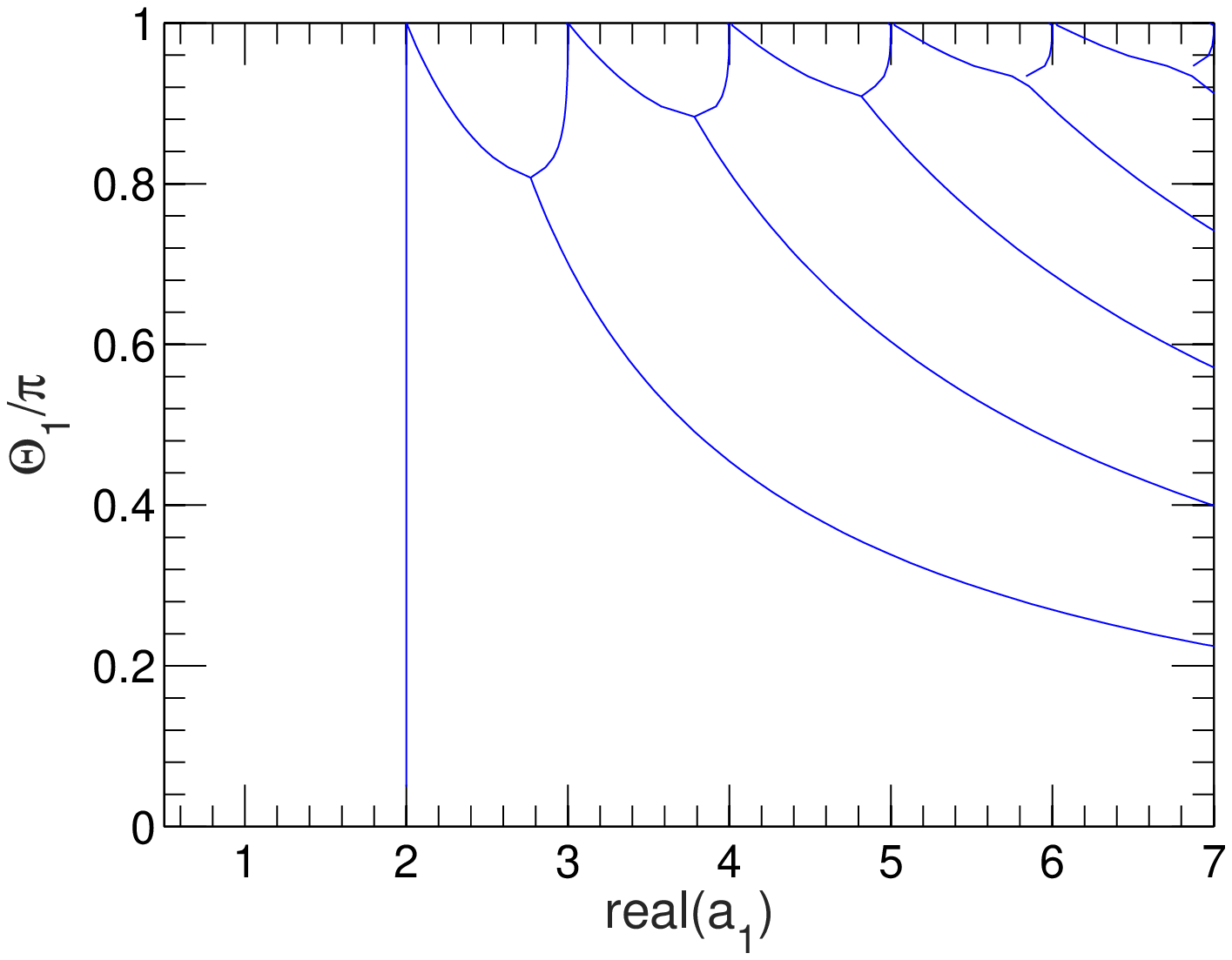}
		 \caption{}
	\end{subfigure}
	 \caption{(\textit{a}) Schematic of the local problem near the contact line pinned at $O$, with surfactant on the interface $(\theta=0)$ between two incompressible fluids with viscosities $\mu_1$ (bottom fluid) and $\mu_2$ (top fluid) and contact angles $\Theta_1$ and $\Theta_2$, respectively. (\textit{b}) Plot of the real part of admissible exponents for the radial dependence of the stream function, calculated from \eqref{eq:ComplexLambda}, against the contact angle $\Theta_1$. This gives the exponents in the asymptotic series capturing the behaviour of the fluid as $r \to 0$. Importantly, this shows no admissible exponents with $1<\text{real}(a_1)<2$, which means that for any contact angle and viscosity ratio, the nature of the dominant singularity presented in the main text for a single-fluid flow with contact angle $\pi/2$ is generic.}
	 \label{fig:Exponents}
\end{figure}

We consider the generalisation of the local analysis of the single-fluid flow with $\pi/2$ contact angle made in \S\ref{sec:corners}. As depicted in figure \ref{fig:Exponents}(\textit{a}), we now assume a two-fluid incompressible Stokes flow with arbitrary viscosities. We assume that the interface is locally flat near the contact lines and remains pinned to the flat walls of the cavity at the contact line location at point $O$, which corresponds to the coordinates $(x,y)=(-1,0)$ in the original problem described in figure \ref{fig:Domain}. We allow the contact angle $\Theta$ to vary between 0 and $\pi$, thereby relaxing the assumption made in the main part of the text. We find local approximations to the stream functions in each fluid, $\psi_{(1)}$ in fluid $1$ (bottom fluid), and $\psi_{(2)}$ in fluid $2$ (top fluid). These fluids have viscosities $\mu_1$ and $\mu_2$, and  contact angles $\Theta_1$ and $\Theta_2 = \pi- \Theta_1$, respectively. 

We use a plane polar coordinate system with $r$ being the radial direction from the origin, and $\theta$ the angular coordinate, with the interface along the line $\theta=0$ (see figure \ref{fig:Exponents}\textit{a}). 
Similar to the model presented in \S\ref{sec:model}, we formulate the flow problem with the stream function, which follows the biharmonic equation \eqref{eq:bih} in each fluid, with no-flux and no-penetration at the cavity wall, and no penetration at the interface, which is assumed fixed and locally flat. At the interface $(r\geq 0,\theta=0)$, we also assume continuity of the tangential velocity, whilst the tangential dynamic boundary condition now becomes, in dimensional form,
\begin{equation}\label{eq:tangentialdynamicBC}
	-{\boldsymbol \tau }\cdot \llbracket \boldsymbol{\sigma}^*\rrbracket \cdot \mathbf{n} = {\boldsymbol \tau }\cdot {\bf \nabla}^*_s \gamma^*
\end{equation} 
where the jump in tangential stress across the interface is balanced by the surfactant-induced Marangoni stress.  The jump bracket is defined as $\llbracket \boldsymbol{\sigma}^*\rrbracket= \boldsymbol{\sigma}_2^*-\boldsymbol{\sigma}_1^*$. The stress tensor $\boldsymbol{\sigma}^*$ is assumed Newtonian for both fluids. The unit  tangential and normal vectors at the interface follow  ${\boldsymbol{\tau}}=(1,0)$ and ${\bf n}=(0,1)$, in polar coordinates, as depicted in figure \ref{fig:Exponents}(\textit{a}). The surface gradient operator is defined as ${\bf \nabla}^*_s={\bf \nabla}^* \cdot(\mathsf{I}-{\bf n} \otimes {\bf n})$, with $\otimes$ the outer product. Similar to before, we non-dimensionalize this problem using \eqref{scales}, taking fluid 1 as the reference fluid, then we linearize the surfactant distribution around $\bar{\Gamma}$, and decompose the perturbation for each variable in the form $\hat{\Gamma}(r)e^{-\lambda t}$ for example. Relating all variables to the streamfunction, as done previously in \S\ref{sec:linearization}, the  tangential dynamic boundary condition \eqref{eq:tangentialdynamicBC} becomes for each mode, in polar coordinates,
\begin{multline}\label{2FluidsPolarBoundCond}
-\alpha\left(\frac{1}{r^2} \psi_{(1)\theta \theta} + \frac{1}{r}\psi_{(1)r} -  \frac{\mu_r}{r^2} \psi_{(2)\theta \theta}  - \frac{\mu_r}{r}\psi_{(2)r} \right)
\\
=\frac{1}{r} \psi_{(1)rr \theta} -\frac{2}{r^2}\psi_{(1)r \theta} + \frac{2}{r^3} \psi_{(1)\theta} =\frac{1}{r} \psi_{(2)rr \theta} -\frac{2}{r^2}\psi_{(2)r \theta} + \frac{2}{r^3} \psi_{(2)\theta},
	\qquad \text{on } \theta = 0.
\end{multline} 
where $\alpha=\lambda/\bar{\Gamma}$ is the decay rate of each mode, and $\mu_r=\mu_2/\mu_1$ is the viscosity ratio between the two fluids.
The first term incorporates the difference in shear stress between the lower and the upper fluid; the middle and final terms describe stretching of the interface.  Continuity of the tangential velocity field along the interface requires both fluids to stretch at an equal rate.

We  seek separable solutions to the above problem such that the leading order terms in each series is $\psi_{(1)} = r^{a_1}f_{a_1}(\theta)$ and $\psi_{(2)} = r^{a_2}f_{a_2}(\theta)$ where the functions  $f$ are given by \eqref{f_1} to \eqref{f_mu odd}.
Applying the boundary conditions we find that the constants in the functions $f_{a_1}$ and $f_{a_2}$ depend on the contact angle, whilst the exponents $a_1$ and $a_2$ satisfy
\begin{equation}\label{eq:conditionf1f2}
    	\left(a_1^2-3a_1 +2\right)f_{a_1}^{\prime}(0;\Theta_1) = 0 
    	\qquad \text{or} \qquad
	\left(a_2^2-3a_2 +2\right)f_{a_2}^{\prime}(0;-\Theta_2) = 0. 
\end{equation} 
The conditions in \eqref{eq:conditionf1f2} can be satisfied with $a_1=a_2=2$, based on the first bracket in each condition, such that the stream function solutions must involve series with  exponent equal to $2$. However, we ask the question whether taking either $f_{a_1}^{\prime}(0;\Theta_1) = 0 $ or $f_{a_2}^{\prime}(0;-\Theta_2) = 0$ in \eqref{eq:conditionf1f2}  can give us an exponent with real part between $1$ and $2$. This is important as an exponent less than $2$ would give us a stronger corner singularity than that discussed in the main text for the case of a single-fluid flow with a contact angle of $\pi/2$. Exponents with real part less than or equal to $1$, from the conditions \eqref{eq:conditionf1f2}, are rejected on physical grounds to avoid the radial velocity to diverge or be non-zero as $r \to 0$.

When we impose the condition $f_{a_1}^{\prime}(0,\Theta_1) = 0$, we find that the exponent must obey the condition which is the same as found in Moffatt's problem for a flow near a corner of angle $\Theta_1$ subject to zero velocity boundary conditions at the boundaries (and similarly for $a_2$) \citep{moffatt1964viscous}. This condition is
\begin{equation}\label{eq:ComplexLambda}
    (a_1-1)\sin{(\Theta_1)} = \pm \sin{(\Theta_1(a_1-1))} .
\end{equation}
By inspection we can see that $a_1=0,1,2$ (and similarly for $a_2$) are solutions of \eqref{eq:ComplexLambda}, for any value of $\Theta_1$, and any integer is a solution when $\Theta_1 = \pi$. We show that a solution with $1 < \text{real}(a_1)<2$ cannot exist for $0\leq \Theta_1 \leq \pi$ in figure \ref{fig:Exponents}\textit{(b)} where we have plotted the locations of the real parts of the admissible solutions of \eqref{eq:ComplexLambda} against $\Theta_1$. We also note that none of the exponents in the expansions for $\psi_{(1)}$ or $\psi_{(2)}$ depend on the viscosity ratio. Hence, in this problem the exponent in the dominant term in both stream functions will be $2$ for any contact angle and viscosity ratio, giving the same type of singularities as presented in the main body of the paper for this broader class of problems.

\bibliographystyle{jfm}
\bibliography{Refs}

\end{document}